\documentclass[10pt]{iopart}

\usepackage{graphicx}            
\usepackage[dvipsnames]{xcolor}  

\begin{document}

\title[2021 Recommended frequency values]{The CIPM list ``Recommended values of standard frequencies'': 2021 update}

\author{H. S. Margolis$^1$, G. Panfilo$^2$, G. Petit$^2$, C. Oates$^3$, T. Ido$^4$ and S. Bize$^5$}

\address{$^1$ National Physical Laboratory, Hampton Road, Teddington, TW11 0LW, United Kingdom}
\address{$^2$ Bureau International des Poids et Mesures, 92310 S\`evres, France}
\address{$^3$ National Institute of Standards and Technology (NIST), Boulder, CO 80305, United States of America}
\address{$^4$ National Institute of Information and Communications Technology, 4-2-1 Nukui-kitamachi, Koganei, Tokyo, 184-8795, Japan}
\address{$^5$ LNE-SYRTE, Observatoire de Paris, Universit\'e PSL, CNRS, Sorbonne Universit\'e, 61 avenue de l’Observatoire, 75014 Paris, France}

\ead{helen.margolis@npl.co.uk}

\vspace{10pt}

\begin{abstract}
This paper gives a detailed account of the analysis underpinning the 2021 update to the list of standard reference frequency values recommended by the International Committee for Weights and Measures (CIPM). This update focused on a subset of atomic transitions that are secondary representations of the second (SRS) or considered as potential SRS. As in previous updates in 2015 and 2017, methods for analysing over-determined data sets were applied to make optimum use of the worldwide body of published clock comparison data. To ensure that these methods were robust, three independent calculations were performed using two different algorithms. The 2021 update differed from previous updates in taking detailed account of correlations among the input data, a step shown to be important in deriving unbiased frequency values and avoiding underestimation of their uncertainties. It also differed in the procedures used to assess input data and to assign uncertainties to the recommended frequency values, with previous practice being adapted to produce a fully consistent output data set consisting of frequency ratio values as well as absolute frequencies. These changes are significant in the context of an anticipated redefinition of the second in terms of an optical transition or transitions, since optical frequency ratio measurements will be critical for verifying the international consistency of optical clocks prior to the redefinition. In the meantime, the reduced uncertainties for optical SRS resulting from this analysis significantly increases the weight that secondary frequency standards based on these transitions can have in the steering of International Atomic Time (TAI).
\end{abstract}

\vspace{2pc}


%
\ioptwocol

\section{Introduction}
\label{sec:Intro}

The historical development of the list of standard reference frequency values recommended by the International Committee for Weights and Measures (CIPM) has been well described by Riehle {\it et al.}~\cite{Riehle2018}. Today it contains recommended values of standard frequencies for applications that include both the practical realisation of the definition of the metre and secondary representations of the definition of the second (SRS). The 2021 update to the list, described in this paper, focused on 14 atomic transitions (table~\ref{tab:Transitions}) which had either already been adopted as SRS, or which were considered to be potential candidates for becoming SRS. The analysis underpinning the update was performed under the auspices of the Working Group on Frequency Standards (WGFS), a joint working group of the Consultative Committee for Length (CCL) and the Consultative Committee for Time and Frequency (CCTF), who are charged by the CIPM with maintaining the list and making proposals for recommendations to the relevant consultative committee~\cite{WGFS_TOR}. The updated recommended frequency values and uncertainties were approved by the CCTF at their 22nd meeting in March 2021~\cite{CCTF_Rec2_2021} and became active on 13th April 2022, following publication on the website of the International Bureau of Weights and Measures (BIPM)~\cite{BIPM_WebList}. 

The list of recommended frequencies plays an important role in progress towards an anticipated redefinition of the SI second based on an optical transition or transitions. Optical frequency standards are already used as secondary frequency standards (SFS), alongside caesium primary frequency standards (PFS), for calibration of the scale interval of International Atomic Time (TAI)~\cite{Petit2015,Panfilo2019,BIPM_PSFSPlot} and the number and frequency of these contributions is expected to increase significantly in the coming years. SFS contribute to TAI using the recommended frequency value and uncertainty of the SRS on which they are based, and so the lower these uncertainties are, the more benefit these contributions will bring to the stability and accuracy of TAI. The analysis underpinning the derivation of the recommended frequency values can also help to verify the consistency of optical frequency ratio measurements, another key prerequisite for a redefinition of the second.

\begin{table}
\caption{\label{tab:Transitions}Atomic transitions included in the 2021 least-squares adjustment.}
\begin{indented}
\item[]\begin{tabular}{@{}lll}
\br
      & Atomic  &                      \\
Label & species & Reference transition \\
\mr
$\nu_1$    & $^{115}$In$^+$ & 5s$^2\;^1$S$_0$--5s5p$\;^3$P$_0$                              \\
$\nu_2$    & $^1$H          & 1s$\;^2$S$_{1/2}$--2s$\;^2$S$_{1/2}$                          \\
$\nu_3$    & $^{199}$Hg     & 6s$^2\;^1$S$_0$--6s6p$\;^3$P$_0$                              \\
$\nu_4$    & $^{27}$Al$^+$  & 3s$^2\;^1$S$_0$--3s3p$\;^3$P$_0$                              \\
$\nu_5$    & $^{199}$Hg$^+$ & 5d$^{10}$6s$\;^2$S$_{1/2}\;(F=0)$--5d$^9$6s$^2\;^2$D$_{5/2}\;(F=2)$         \\
$\nu_6$    & $^{171}$Yb$^+$ & 6s$\;^2$S$_{1/2}\;(F=0)$--5d$\;^2$D$_{3/2}\;(F=2)$            \\
$\nu_7$    & $^{171}$Yb$^+$ & 6s$\;^2$S$_{1/2}\;(F=0)$--4f$^{13}$6s$^2\;^2$F$_{7/2}\;(F=3)$ \\
$\nu_8$    & $^{171}$Yb     & 6s$^2\;^1$S$_0$--6s6p$\;^3$P$_0$                              \\
$\nu_9$    & $^{40}$Ca      & 4s$^2\;^1$S$_0$--4s4p$\;^3$P$_1$                              \\
$\nu_{10}$ & $^{88}$Sr$^+$  & 5s$\;^2$S$_{1/2}$--4d$\;^2$D$_{5/2}$                          \\
$\nu_{11}$ & $^{88}$Sr      & 5s$^2\;^1$S$_0$--5s5p$\;^3$P$_0$                              \\
$\nu_{12}$ & $^{87}$Sr      & 5s$^2\;^1$S$_0$--5s5p$\;^3$P$_0$                              \\
$\nu_{13}$ & $^{40}$Ca$^+$  & 4s$\;^2$S$_{1/2}$--3d$\;^2$D$_{5/2}$                          \\
$\nu_{14}$ & $^{87}$Rb      & 5s$\;^2$S$_{1/2}\;(F=1)$--5s$^2\;^2$S$_{1/2}\;(F=2)$          \\
$\nu_{15}$ & $^{133}$Cs     & 6s$\;^2$S$_{1/2}\;(F=3)$--6s$^2\;^2$S$_{1/2}\;(F=4)$          \\
\br
\end{tabular}
\end{indented}
\end{table}

Ratios between unperturbed atomic transition frequencies are dimensionless quantities given by nature. For a collection of frequency standards based on $N_{\rm S}$ different reference transitions with frequencies $\nu_k$ ($k=1,2,\ldots ,N_{\rm S}$), it is in principle possible to measure a total of $N_{\rm S}(N_{\rm S}-1)/2 $ different frequency ratios, but only $N_{\rm S}-1$ of these are independent. 
Since 2015, the worldwide body of clock comparison data available to the WGFS has contained an increasing number of optical frequency ratio measurements, in addition to absolute frequency measurements relative to caesium primary frequency standards.
This means that the clock comparison data set is over-determined, i.e. it is possible to derive some frequency ratios from the results of several different experiments. 
To make full use of the available data, new analysis methods were developed~\cite{Margolis2015} and were applied for the first time to update the list of recommended values of standard frequencies in 2015. By the time of the next update in 2017, a second method was also available~\cite{Robertsson2016}, providing important verification of the results. These methods were employed again in the 2021 update to the list described here, but with several significant differences:
\begin{enumerate}
\item Correlations among the input data were given detailed consideration and taken into account in the analysis. As previously pointed out in references~\cite{Margolis2015} and \cite{Margolis2016}, this is critical to ensure that the recommended frequency values are unbiased and that their uncertainties are properly estimated.
\item Reflecting the increasing importance of optical frequency ratios for verifying the international consistency of optical clocks as progress is made towards a redefinition of the second, these were computed and are provided as an appendix to the list of recommended frequency values.
\item In assessing the input data and considering the uncertainty to be assigned to each recommended frequency value, the rules and criteria set out in~\cite{Riehle2018} were modified to ensure that the output data set, which consists of correlated absolute frequency measurements and frequency ratio measurements, is internally self-consistent.
\end{enumerate}

The inclusion of correlations in the analysis, combined with the larger number of new measurements compared to previous updates, significantly increased the effort involved in computing the new recommended frequency values. We note that the work was done under strict time constraints, imposed by the date of the CCTF meeting in March 2021. Some choices therefore had to be made in the analysis that might, in retrospect, have differed in some respects had more time been available for extended interactions with the groups who made the measurements used as input.

The paper is organised as follows. A brief description of the analysis methods is given in section~\ref{sec:AnalysisMethods}, while the new measurement data available for the 2021 update is presented in section~\ref{sec:NewData}. Section~\ref{sec:Correlations} describes the approach that was taken to compute correlation coefficients between the input data. In section~\ref{sec:InputMods} we summarise all modifications that were made to the input data, and describe the rationale for these. The recommended values that result from the analysis are presented and discussed in section~\ref{sec:Results}. We finish in section~\ref{sec:Conclusion} with some conclusions and perspectives that may be relevant for future discussions within the WGFS and the CCTF, in the period leading up to an anticipated optical redefinition of the SI second.

\section{Analysis methods}
\label{sec:AnalysisMethods}

To ensure that the methods used to derive optimal values for the frequencies $\nu_k$ are robust, three independent calculations were carried out based on two different algorithms.

\subsection{Algorithm 1}

The first approach used is a least-squares adjustment based on the well-established method employed by the CODATA Task Group on Fundamental Constants to derive a self-consistent set of values for the fundamental physical constants~\cite{Mohr2000}. The method is described in detail in~\cite{Margolis2015} and hence is outlined only briefly here. 

The input data to the least-squares adjustment are a set of $N$ frequency ratio measurements $q_i$, together with their standard uncertainties $u_i$ and correlation coefficients $r(q_i,q_j)$, from which their variances $u_i^2$ and covariances $u_{ij}=u_{ji}$ are computed. The input data includes optical frequency ratios, microwave frequency ratios and optical-microwave frequency ratios, but all are handled in a similar way, with absolute frequency measurements treated simply as a special case of frequency ratios involving a caesium primary frequency standard. 

The measured frequency ratios are expressed as a function of one or more of a set of $M=N_{\rm S}-1$ independent adjusted frequency ratios $z_j$, yielding a set of $N$ equations. It is the values of these adjusted frequency ratios that are optimized in the least squares adjustment.

In most cases, the equations relating the measured frequency ratios to the adjusted frequency ratios are nonlinear. So that linear matrix methods can be employed, the equations are therefore linearised prior to the least-squares adjustment by using a Taylor expansion around initial estimates of the adjusted frequency ratios. 
This linear approximation means that the best estimates for the values of the adjusted frequency ratios (together with their variances and covariances) are not exact solutions to the original nonlinear equations, but the values obtained from the least-squares adjustment are used as starting values for a new linear approximation and another least-squares adjustment performed. This process is repeated until the values of the adjusted frequency ratios converge. Any other frequency ratio of interest (and its uncertainty) can then be calculated from the adjusted frequency ratios and their covariance matrix.

Two independent implementations of this algorithm were used for the computation, one written in MATLAB\textsuperscript{\textregistered} and the other written in Mathematica\textsuperscript{\textregistered}.  

\subsection{Algorithm 2}

The second algorithm used to derive optimal values for the frequencies $\nu_{k}$ uses a different conceptual approach described in~\cite{Robertsson2016}, based on the examination of closed loops in a graph theory framework. By analysing figure~\ref{fig:InputData}, a number of closed loops can easily be identified. As in the first algorithm, the equations reporting the ratios are nonlinear, so to simplify the treatment the logarithms of the frequencies are used, which converts the problem to a linear least squares one. The logarithms of all frequency ratios in each closed loop should add up to zero. This provides a set of conditions that are used in a Lagrange multiplier method to identify the basis vectors for the residual space in the least squares calculation. A projection on this subspace gives the corrections to the experimental ratio values.

A single implementation of this algorithm, written in Matlab, was used for the computation.

\subsection{Numerical precision}

In all cases, due to the extremely high accuracy with which frequency ratios have been measured, care had to be taken to perform numerical calculations to sufficiently high precision (more than 18 significant figures). This was achieved using routines designed for high precision floating point arithmetic, the standard double-precision floating point format being insufficient.

\section{New measurement data since 2017}
\label{sec:NewData}

\begin{table*}
\caption{\label{tab:2017_Absolutes}Absolute frequency measurement data used in the 2017 least-squares adjustment.  All frequency values and uncertainties listed are the published ones, before adjustments made by the WGFS (see section~\ref{sec:InputMods}).}
\begin{indented}
\item[]\begin{tabular}{@{}llllll}
\br
Clock transition & Atomic species & Measured frequency / Hz & Fractional uncertainty & Reference & Label  \\
\mr
$\nu_1$ & $^{115}$In$^+$ & 1267\,402\,452\,899\,920(230)   & $1.8\times 10^{-13}$ & \cite{vonZanthier2000} & $q_1$ \\
        &     & 1267\,402\,452\,901\,265(256)   & $2.0\times 10^{-13}$ & \cite{Wang2007a}$^{\rm a}$ & $q_2$ \\
        &     & 1267\,402\,452\,901\,049.9(6.9) & $5.9\times 10^{-15}$ & \cite{Ohtsubo2017} & $q_3$  \\
\ms 
$\nu_2$ & $^1$H & 1233\,030\,706\,593\,517.5(5.0)  & $4.1\times 10^{-15}$ & \cite{Parthey2011} & $q_{4}$ \\
        &       & 1233\,030\,706\,593\,509.0(5.5)  & $4.5\times 10^{-15}$ & \cite{Matveev2013} & $q_{5}$ \\
\ms 
$\nu_3$ & $^{199}$Hg & 1128\,575\,290\,808\,155.1(6.4)  & $5.7\times 10^{-15}$ & \cite{McFerran2012,McFerran2015}$^{\rm b}$ & $q_{6}$ \\
        &            & 1128\,575\,290\,808\,154.62(0.41) & $3.6\times 10^{-16}$ & \cite{Tyumenev2016}              & $q_{7}$ \\
\ms 
$\nu_4$ & $^{27}$Al$^+$  & 1121\,015\,393\,207\,851(6)       & $5.4\times 10^{-15}$ & \cite{Rosenband2007} & $q_{8}$ \\
\ms 
$\nu_5$ & $^{199}$Hg$^+$ & 1064\,721\,609\,899\,145.30(0.69) & $6.5\times 10^{-16}$ & \cite{Stalnaker2007} & $q_{9}$ \\
\ms 
$\nu_6$ & $^{171}$Yb$^+$ E2 & {\lineup\0}688\,358\,979\,309\,308.0(2.14)  & $3.1\times 10^{-15}$ & \cite{12thCCL}     & $q_{10}$ \\
        &                   & {\lineup\0}688\,358\,979\,309\,306.97(0.73) & $1.1\times 10^{-15}$ & \cite{Tamm2009}$^{\rm c}$    & $q_{11}$ \\
        &                   & {\lineup\0}688\,358\,979\,309\,310(9)       & $1.3\times 10^{-14}$ & \cite{Webster2010} & $q_{12}$ \\
        &                   & {\lineup\0}688\,358\,979\,309\,307.82(0.36) & $5.2\times 10^{-16}$ & \cite{Tamm2014}    & $q_{13}$ \\
        &                   & {\lineup\0}688\,358\,979\,309\,308.42(0.42) & $6.1\times 10^{-16}$ & \cite{Godun2014}    & $q_{14}$ \\
\ms 
$\nu_7$ & $^{171}$Yb$^+$ E3 & {\lineup\0}642\,121\,496\,772\,657(12) & $1.9\times 10^{-14}$ & \cite{Hosaka2009} & $q_{15}$ \\
        &      & {\lineup\0}642\,121\,496\,772\,645.15(0.52) & $8.1\times 10^{-16}$ & \cite{Huntemann2012} & $q_{16}$ \\
        &      & {\lineup\0}642\,121\,496\,772\,646.22(0.67) & $1.0\times 10^{-15}$ & \cite{King2012}      & $q_{17}$ \\
        &      & {\lineup\0}642\,121\,496\,772\,644.91(0.37) & $5.8\times 10^{-16}$ & \cite{Godun2014}     & $q_{18}$ \\
        &      & {\lineup\0}642\,121\,496\,772\,645.36(0.25) & $3.9\times 10^{-16}$ & \cite{Huntemann2014} & $q_{19}$ \\
\ms 
$\nu_8$ & $^{171}$Yb & {\lineup\0}518\,295\,836\,590\,864(28)      & $5.4\times 10^{-14}$ & \cite{Kohno2009}     & $q_{20}$ \\
        &            & {\lineup\0}518\,295\,836\,590\,863.1(2.0)   & $3.9\times 10^{-15}$ & \cite{Yasuda2012}    & $q_{21}$ \\
        &            & {\lineup\0}518\,295\,836\,590\,865.2(0.7)   & $1.4\times 10^{-15}$ & \cite{Lemke2009}     & $q_{22}$ \\
        &            & {\lineup\0}518\,295\,836\,590\,863.5(8.1)   & $1.6\times 10^{-14}$ & \cite{Park2013}      & $q_{23}$ \\
        &            & {\lineup\0}518\,295\,836\,590\,863.59(0.31) & $6.0\times 10^{-16}$ & \cite{Pizzocaro2017} & $q_{24}$ \\
        &            & {\lineup\0}518\,295\,836\,590\,863.38(0.57) & $1.1\times 10^{-15}$ & \cite{Kim2017}       & $q_{25}$ \\
\ms 
$\nu_9$ & $^{40}$Ca  & {\lineup\0}455\,986\,240\,494\,144.0(5.3) & $1.2\times 10^{-14}$ & \cite{Degenhardt2005} & $q_{26}$ \\
        &            & {\lineup\0}455\,986\,240\,494\,135.8(3.4) & $7.5\times 10^{-15}$ & \cite{Wilpers2006,Wilpers2007} & $q_{27}$ \\
\ms 
$\nu_{10}$ & $^{88}$Sr$^+$ & {\lineup\0}444\,779\,044\,095\,484.6(1.5) & $3.4\times 10^{-15}$ & \cite{Margolis2004} & $q_{28}$ \\
           &               & {\lineup\0}444\,779\,044\,095\,484(15)    & $3.4\times 10^{-14}$ & \cite{Dube2005}        & $q_{29}$ \\
           &               & {\lineup\0}444\,779\,044\,095\,485.5(0.9)  & $2.0\times 10^{-15}$ & \cite{Madej2012}    & $q_{30}$ \\
           &               & {\lineup\0}444\,779\,044\,095\,486.71(0.24) & $5.3\times 10^{-16}$ & \cite{Barwood2014} & $q_{31}$ \\
           &               & {\lineup\0}444\,779\,044\,095\,485.27(0.75) & $1.7\times 10^{-15}$ & \cite{Dube2017}    & $q_{32}$ \\
\ms 
$\nu_{11}$ & $^{88}$Sr & {\lineup\0}429\,228\,066\,418\,009(32) & $7.5\times 10^{-14}$ & \cite{Baillard2007}  & $q_{33}$ \\
        &        & {\lineup\0}429\,228\,066\,418\,008.3(2.1)   & $4.9\times 10^{-15}$ & \cite{Morzynski2015} & $q_{34}$ \\
        &        & {\lineup\0}429\,228\,066\,418\,007.3(2.9)   & $6.8\times 10^{-15}$ & \cite{Morzynski2015} & $q_{35}$ \\
\ms 
$\nu_{12}$ & $^{87}$Sr & {\lineup\0}429\,228\,004\,229\,874.0(1.1)   & $2.6\times 10^{-15}$ & \cite{Boyd2007}      & $q_{36}$ \\
           &           & {\lineup\0}429\,228\,004\,229\,873.65(0.37) & $8.6\times 10^{-16}$ & \cite{Campbell2008}  & $q_{37}$ \\
           &           & {\lineup\0}429\,228\,004\,229\,873.6(1.1)   & $2.6\times 10^{-15}$ & \cite{Baillard2008}  & $q_{38}$ \\
           &           & {\lineup\0}429\,228\,004\,229\,874.1(2.4)   & $5.6\times 10^{-15}$ & \cite{Baillard2008}  & $q_{39}$ \\
           &           & {\lineup\0}429\,228\,004\,229\,872.9(0.5)   & $1.2\times 10^{-15}$ & \cite{Falke2011}     & $q_{40}$ \\
           &           & {\lineup\0}429\,228\,004\,229\,873.9(1.4)   & $3.3\times 10^{-15}$ & \cite{Yamaguchi2012} & $q_{41}$ \\
           &           & {\lineup\0}429\,228\,004\,229\,872.0(1.6)   & $3.7\times 10^{-15}$ & \cite{Akamatsu2014b} & $q_{42}$ \\
           &           & {\lineup\0}429\,228\,004\,229\,873.56(0.49) & $1.1\times 10^{-15}$ & \cite{Tanabe2015}    & $q_{43}$ \\
           &           & {\lineup\0}429\,228\,004\,229\,873.7(1.4)   & $3.3\times 10^{-15}$ & \cite{Lin2015}       & $q_{44}$ \\
           &           & {\lineup\0}429\,228\,004\,229\,873.13(0.17) & $4.0\times 10^{-16}$ & \cite{Falke2014}     & $q_{45}$ \\
           &           & {\lineup\0}429\,228\,004\,229\,873.10(0.13) & $3.1\times 10^{-16}$ & \cite{LeTargat2013}  & $q_{46}$ \\
           &           & {\lineup\0}429\,228\,004\,229\,872.92(0.12) & $2.8\times 10^{-16}$ & \cite{Lodewyck2016}  & $q_{47}$ \\
           &           & {\lineup\0}429\,228\,004\,229\,872.97(0.16) & $3.7\times 10^{-16}$ & \cite{Grebing2016}   & $q_{48}$ \\
           &           & {\lineup\0}429\,228\,004\,229\,873.04(0.11) & $2.6\times 10^{-16}$ & \cite{Grebing2016}   & $q_{49}$ \\
           &           & {\lineup\0}429\,228\,004\,229\,872.97(0.40) & $9.3\times 10^{-16}$ & \cite{Hachisu2017a}  & $q_{50}$ \\
           &           & {\lineup\0}429\,228\,004\,229\,872.99(0.18) & $4.3\times 10^{-16}$ & \cite{Hachisu2017b}  & $q_{51}$ \\
\ms 
$\nu_{13}$ & $^{40}$Ca$^+$ & {\lineup\0}411\,042\,129\,776\,393.2(1.0)   & $2.4\times 10^{-15}$ & \cite{Chwalla2009}   & $q_{52}$ \\
           &               & {\lineup\0}411\,042\,129\,776\,398.4(1.2)   & $2.9\times 10^{-15}$ & \cite{Matsubara2012} & $q_{53}$ \\
           &               & {\lineup\0}411\,042\,129\,776\,400.5(1.2)   & $2.9\times 10^{-15}$ & \cite{Huang2016}     & $q_{54}$ \\
           &               & {\lineup\0}411\,042\,129\,776\,401.7(1.1)   & $2.7\times 10^{-15}$ & \cite{Huang2016}     & $q_{55}$ \\
\ms 
$\nu_{14}$ & $^{87}$Rb & \lineup\0\0\0\0\0\0 6834\,682\,610.904\,3129($3.0\times 10^{-6}$) & $4.4\times 10^{-16}$ & \cite{SYRTETAIData}    & $q_{56}$ \\
           &           & \lineup\0\0\0\0\0\0 6834\,682\,610.904\,3070($3.1\times 10^{-6}$) & $4.5\times 10^{-16}$ & \cite{Ovchinnikov2015} & $q_{57}$ \\
           &           & \lineup\0\0\0\0\0\0 6834\,682\,610.904\,3125($2.1\times 10^{-6}$) & $3.1\times 10^{-16}$ & \cite{Guena2017}       & $q_{58}$ \\
\br
\end{tabular}
\item[] $^{\rm a}$Another published absolute frequency measurement of $\nu_1$~\cite{Wang2007b} was omitted from the input data set, because it is based on data that is apparently identical to that used in~\cite{Wang2007a}, but reports a much lower uncertainty of 63 Hz without explanation being provided. 
\item[] $^{\rm b}$The value of $q_6$ is the corrected value given in~\cite{McFerran2015}, originating from the experiment described in~\cite{McFerran2012}. 
\item[] $^{\rm c}$The value of $q_{11}$ is the value published in~\cite{Tamm2009}, but corrected for the blackbody radiation shift.
\end{indented}
\end{table*}

\begin{table*}
\caption{\label{tab:2017_Ratios}Frequency ratio measurement data used in the 2017 least-squares adjustment.}
\begin{indented}
\item[]\begin{tabular}{@{}llllll}
\br
Frequency  &                 &                 & Fractional  &           &        \\
ratio      & Atomic species  &  Measured ratio & uncertainty & Reference & Label  \\
\mr
$\nu_3/\nu_{12}$ & $^{199}$Hg/$\,^{87}$Sr & {\lineup\0\0\0\0\0\,}2.629\,314\,209\,898\,909\,60(22) & $8.4\times 10^{-17}$ & \cite{Yamanaka2015} & $q_{59}$ \\
                 &                            & {\lineup\0\0\0\0\0\,}2.629\,314\,209\,898\,909\,15(46) & $1.7\times 10^{-16}$ & \cite{Tyumenev2016} & $q_{60}$ \\
\ms 
$\nu_3/\nu_{14}$ & $^{199}$Hg/$\,^{87}$Rb & 165\,124.754\,879\,997\,258(62)  & $3.8\times 10^{-16}$ & \cite{Tyumenev2016} & $q_{61}$ \\
\ms
$\nu_4/\nu_5$ & $^{27}$Al$^+$/$\,^{199}$Hg$^+$ & {\lineup\0\0\0\0\0\,}1.052\,871\,833\,148\,990\,438(55) &  $5.2\times 10^{-17}$ & \cite{Rosenband2008} & $q_{62}$ \\
\ms
$\nu_6/\nu_7$ & $^{171}$Yb$^+$ E2$\,$/$\,^{171}$Yb$^+$ E3 & {\lineup\0\0\0\0\0\,}1.072\,007\,373\,634\,206\,30(36) & $3.4\times 10^{-16}$ & \cite{Godun2014} & $q_{63}$ \\
\ms 
$\nu_8/\nu_{12}$ & $^{171}$Yb /$\,^{87}$Sr & {\lineup\0\0\0\0\0\,}1.207\,507\,039\,343\,3412(17)   & $1.4\times 10^{-15}$ & \cite{Akamatsu2014a,Akamatsu2014aErr}$^{\rm a}$ & $q_{64}$  \\
 &                        & {\lineup\0\0\0\0\0\,}1.207\,507\,039\,343\,337\,76(29)  & $2.4\times 10^{-16}$ & \cite{Takamoto2015}                   & $q_{65}$ \\
 &                        & {\lineup\0\0\0\0\0\,}1.207\,507\,039\,343\,337\,749(55) & $4.6\times 10^{-17}$ & \cite{Nemitz2016}                     & $q_{66}$ \\
\ms 
$\nu_{11}/\nu_{12}$ & $^{88}$Sr /$\,^{87}$Sr & {\lineup\0\0\0\0\0\,}1.000\,000\,144\,883\,682\,777(23) & $2.3\times 10^{-17}$ & \cite{Takano2017} & $q_{67}$ \\ 
\ms 
$\nu_{12}/\nu_{13}$ & $^{87}$Sr /$\,^{40}$Ca$^+$ & {\lineup\0\0\0\0\0\,}1.044\,243\,334\,529\,6416(25) & $2.4\times 10^{-15}$ & \cite{Matsubara2012} & $q_{68}$ \\
\ms 
$\nu_{12}/\nu_{14}$ & $^{87}$Sr /$\,^{87}$Rb  & {\lineup\0}62\,801.453\,800\,512\,435(21)    & $3.3\times 10^{-16}$ & \cite{Lodewyck2016}  & $q_{69}$ \\
\br
\end{tabular}
\item[] $^{\rm a}$The value $q_{64}$ is the corrected value given in~\cite{Akamatsu2014aErr}, originating from the experiment described in~\cite{Akamatsu2014a}.
\end{indented}
\end{table*}

The starting point for the collection of the input data was the input data file from the previous adjustment performed by the WGFS in 2017. That file included 69 input data points, of which 58 were absolute frequency measurements (table~\ref{tab:2017_Absolutes}). The remaining 11 were frequency ratio measurements that did not involve Cs (table~\ref{tab:2017_Ratios}), most of them optical frequency ratio measurements, but two being optical-microwave frequency ratios. 

\begin{table*}
\caption{\label{tab:New_Absolutes}New absolute frequency measurements included in the 2021 least-squares adjustment. All uncertainties listed are the published ones, before adjustments made by the WGFS (see section~\ref{sec:InputMods}).}
\begin{indented}
\item[]\begin{tabular}{@{}llllll}
\br
Clock transition & Atomic species & Measured frequency / Hz & Fractional uncertainty & Reference & Label \\
\mr
$\nu_1$ & $^{115}$In$^+$    & 1267\,402\,452\,901\,040.1(1.1)   & $8.5\times 10^{-16}$ & \cite{Ohtsubo2020}  & $q_{74}$ \\
\ms
$\nu_4$ & $^{27}$Al$^+$     & 1121\,015\,393\,207\,859.50(0.36) & $3.2\times 10^{-16}$ & \cite{Leopardi2021} & $q_{97}$ \\
\ms
$\nu_7$ & $^{171}$Yb$^+$ E3 & {\lineup\0}642\,121\,496\,772\,645.14(0.26) & $4.0\times 10^{-16}$ & \cite{Baynham2018} & $q_{71}$ \\
        &                   & {\lineup\0}642\,121\,496\,772\,645.10(0.08) & $1.3\times 10^{-16}$ & \cite{Lange2021}   & $q_{98}$ \\
\ms
$\nu_8$    & $^{171}$Yb & {\lineup\0}518\,295\,836\,590\,863.30(0.38) & $7.3\times 10^{-16}$ & \cite{Luo2020}       & $q_{70}$ \\
           &            & {\lineup\0}518\,295\,836\,590\,863.71(0.11) & $2.1\times 10^{-16}$ & \cite{McGrew2019}    & $q_{75}$ \\
           &            & {\lineup\0}518\,295\,836\,590\,863.61(0.13) & $2.6\times 10^{-16}$ & \cite{Pizzocaro2020} & $q_{76}$ \\
           &            & {\lineup\0}518\,295\,836\,590\,863.54(0.26) & $5.0\times 10^{-16}$ & \cite{Kobayashi2020} & $q_{89}$ \\
\ms
$\nu_{12}$ & $^{87}$Sr & {\lineup\0}429\,228\,004\,229\,873.1(0.5)     & $1.0\times 10^{-15}$ & \cite{Hobson2020}   & $q_{72}$ \\
           &           & {\lineup\0}429\,228\,004\,229\,873.00(0.07)   & $1.5\times 10^{-16}$ & \cite{Schwarz2020}  & $q_{73}$ \\
           &           & {\lineup\0}429\,228\,004\,229\,873.082(0.076) & $1.8\times 10^{-16}$ & \cite{Nemitz2021}   & $q_{90}$ \\
           &           & {\lineup\0}429\,228\,004\,229\,873.13(0.40)   & $9.3\times 10^{-16}$ & \cite{Grotti2018}   & $q_{91}$ \\
           &           & {\lineup\0}429\,228\,004\,229\,873.19(0.15)   & $3.5\times 10^{-16}$ & \cite{Leopardi2021} & $q_{96}$ \\
\ms
$\nu_{13}$ & $^{40}$Ca$^+$ & {\lineup\0}411\,042\,129\,776\,400.41(0.23) & $5.6\times 10^{-16}$ & \cite{Huang2020} & $q_{88}$ \\
           &               & {\lineup\0}411\,042\,129\,776\,400.6(0.5)   & $1.2\times 10^{-15}$ & \cite{Huang2020} & $q_{105}$ \\
\br
\end{tabular}
\end{indented}
\end{table*}

\begin{table*}
\caption{\label{tab:New_Ratios}New frequency ratio measurements included in the 2021 least-squares adjustment. All uncertainties listed are the published ones, before adjustments made by the WGFS (see section~\ref{sec:InputMods}).}
\begin{indented}
\item[]\begin{tabular}{@{}llllll}
\br
Frequency  &   &   & Fractional &  &  \\
ratio & Atomic species  &  Measured ratio & uncertainty & Reference & Label  \\
\mr
$\nu_1 / \nu_{12}$ & $^{115}$In$^+$/$\,^{87}$Sr            & {\lineup\0\0\0\0\,}2.952\,748\,749\,874\,8633(23) & $7.7\times 10^{-16}$ & \cite{Ohtsubo2020}   & $q_{78}$ \\
\ms
$\nu_3 / \nu_8$ & $^{199}$Hg/$\,^{171}$Yb           & {\lineup\0\0\0\0\,}2.177\,473\,194\,134\,565\,07(19) & $8.8\times 10^{-17}$  & \cite{Ohmae2020}     & $q_{79}$ \\
\ms
$\nu_4 / \nu_8$ & $^{27}$Al$^+$/$\,^{171}$Yb            & {\lineup\0\0\0\0\,}2.162\,887\,127\,516\,663\,703(13) & $5.9\times 10^{-18}$ & \cite{BACON2021}     & $q_{104}$ \\
\ms
$\nu_4 / \nu_{12}$ & $^{27}$Al$^+$/$\,^{87}$Sr             & {\lineup\0\0\0\0\,}2.611\,701\,431\,781\,463\,025(21)   & $8.0\times 10^{-18}$ & \cite{BACON2021}     & $q_{103}$ \\
\ms
$\nu_6 / \nu_7$ & $^{171}$Yb$^+$ E2$\,$/$\,^{171}$Yb$^+$ E3 & {\lineup\0\0\0\0\,}1.072\,007\,373\,634\,205\,469(37)   & $3.5\times 10^{-17}$ & \cite{Lange2021}$^{\rm a}$     & $q_{99}$ \\
\ms
$\nu_7 / \nu_{12}$ & $^{171}$Yb$^+$ E3$\,$/$\,^{87}$Sr         & {\lineup\0\0\0\0\,}1.495\,991\,618\,544\,900\,976(494)  & $3.3\times 10^{-16}$ &  \cite{Riedel2020}    & $q_{84}$ \\
 &                                       & {\lineup\0\0\0\0\,}1.495\,991\,618\,544\,901\,113(404)  & $2.7\times 10^{-16}$ & \cite{Riedel2020}    & $q_{85}$ \\
 &                                       & {\lineup\0\0\0\0\,}1.495\,991\,618\,544\,900\,644(524)  & $3.5\times 10^{-16}$ & \cite{Riedel2020}    & $q_{86}$ \\
 &                                       & {\lineup\0\0\0\0\,}1.495\,991\,618\,544\,900\,858(299)  & $2.0\times 10^{-16}$ & \cite{Riedel2020}    & $q_{87}$ \\
 &                                       & {\lineup\0\0\0\0\,}1.495\,991\,618\,544\,900\,537(38)   & $2.5\times 10^{-17}$ & \cite{Dorscher2021}  & $q_{92}$ \\
 &                                       & {\lineup\0\0\0\0\,}1.495\,991\,618\,544\,900\,840(344)  & $2.3\times 10^{-16}$ & \cite{Riedel2020}    & $q_{100}$ \\
 &                                       & {\lineup\0\0\0\0\,}1.495\,991\,618\,544\,900\,459(404)  & $2.7\times 10^{-16}$ & \cite{Riedel2020}    & $q_{101}$ \\
\ms
$\nu_8 / \nu_{12}$ & $^{171}$Yb$\,$/$\,^{87}$Sr                & {\lineup\0\0\0\0\,}1.207\,507\,039\,343\,337\,90(70)   & $5.8\times 10^{-16}$  & \cite{Fujieda2018}   & $q_{80}$ \\
 &                                       & {\lineup\0\0\0\0\,}1.207\,507\,039\,343\,338\,41(34)    & $2.8\times 10^{-16}$ & \cite{Grotti2018}    & $q_{81}$ \\
 &                                       & {\lineup\0\0\0\0\,}1.207\,507\,039\,343\,338\,05(34)   & $2.8\times 10^{-16}$  & \cite{Pizzocaro2021} & $q_{82}$ \\
 &                                       & {\lineup\0\0\0\0\,}1.207\,507\,039\,343\,337\,38(30)   & $2.5\times 10^{-16}$  & \cite{Pizzocaro2021} & $q_{83}$ \\
 &                                       & {\lineup\0\0\0\0\,}1.207\,507\,039\,343\,338\,58(49)   & $4.1\times 10^{-16}$  & \cite{Hisai2021}     & $q_{93}$ \\
 &                                       & {\lineup\0\0\0\0\,}1.207\,507\,039\,343\,337\,82(75)   &  $6.2\times 10^{-16}$ & \cite{Kobayashi2020} & $q_{94}$ \\
 &                                       & {\lineup\0\0\0\0\,}1.207\,507\,039\,343\,337\,8482(82) & $6.8\times 10^{-18}$ & \cite{BACON2021}     & $q_{102}$ \\
\ms
$\nu_8 / \nu_{14}$ & $^{171}$Yb$\,$/$\,^{87}$Rb                & 75\,833.197\,545\,114\,174(42) &  $5.5\times 10^{-16}$     & \cite{Kobayashi2020} & $q_{95}$ \\
 &                                       & 75\,833.197\,545\,114\,192(33)      & $4.4\times 10^{-16}$ & \cite{McGrew2019}    & $q_{106}$ \\
\ms
$\nu_{11} / \nu_{12}$ & $^{88}$Sr$\,$/$\,^{87}$Sr                 & {\lineup\0\0\0\0\,}1.000\,000\,144\,883\,682\,831(28)   & $2.8\times 10^{-17}$ & \cite{Origlia2018}   & $q_{77}$ \\
\br
\end{tabular}
\item[] $^{\rm a}$The value reported here is the inverse of the one published in \cite{Lange2021}.
\end{indented}
\end{table*}

\begin{figure}
\includegraphics[width=\columnwidth]{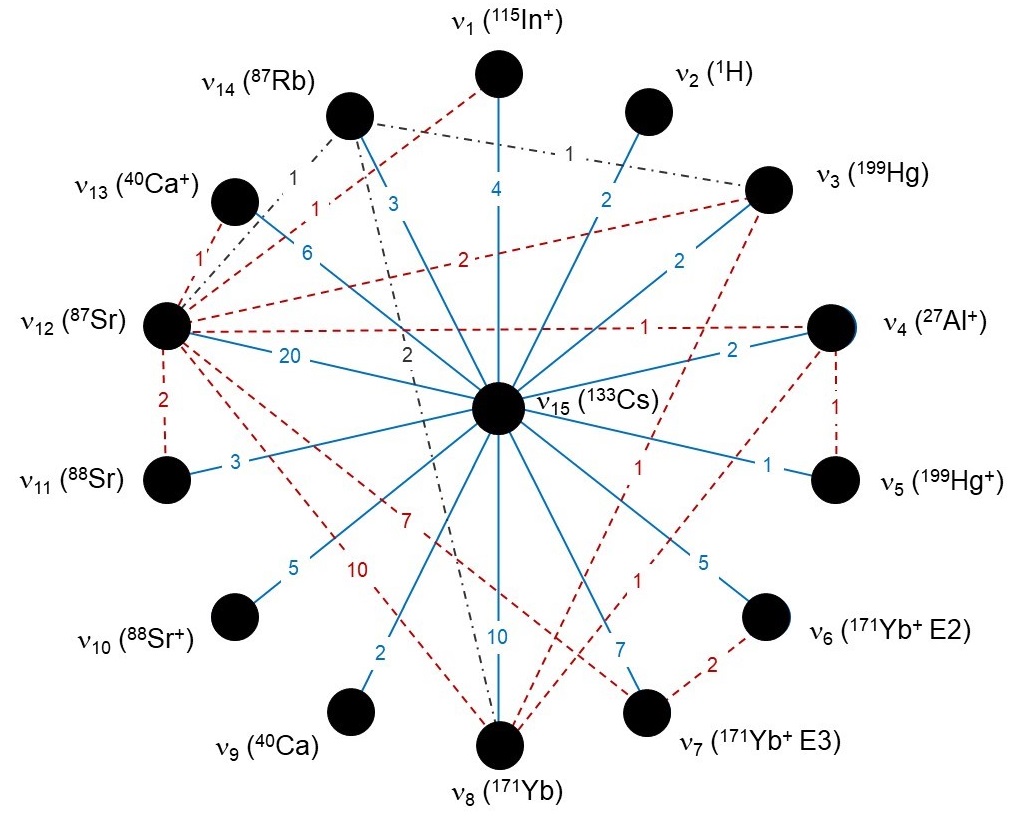}
\caption{\label{fig:InputData}Input data to the 2021 least-squares adjustment. Absolute frequency measurements are indicated by blue solid lines, optical-microwave frequency ratios not involving caesium primary standards by grey dashed-dotted lines, and optical frequency ratio measurements by red dashed lines.
The numbers on each line indicate how many measurements were available in each case.}
\end{figure}

The additional data since 2017 consisted of 15 new absolute frequency measurements (table~\ref{tab:New_Absolutes}) and 22 new frequency ratio measurements (table~\ref{tab:New_Ratios}), all of which met the WGFS requirement of being published in a peer-reviewed journal by March 2021.
However, one of the new frequency measurements ($q_{90}$) included the contribution from the data of $q_{51}$, and so $q_{51}$ was (effectively) excluded from the calculation by multiplying its uncertainty by $10^6$ in the input data file. 
The input data to the 2021 least-squares adjustment thus includes a total of 105 clock comparison measurements, which break down into 72 absolute frequency measurements and 33 frequency ratio measurements not involving Cs. A pictorial representation of the complete body of data is shown in figure~\ref{fig:InputData}.

No new atomic transitions were involved in 2021 as compared to 2017, but some of the new frequency ratios reported had never previously been measured directly. It was noted that a majority of new absolute frequency measurements had been made using International Atomic Time (TAI) as a reference, with fractional uncertainties reaching the low parts in $10^{16}$ level. However the absolute frequency measurement with lowest reported fractional uncertainty ($1.3\times 10^{-16}$) was performed against local caesium fountain primary frequency standards~\cite{Lange2021}. In total, seven absolute frequency measurements had reported fractional uncertainties below $3\times 10^{-16}$, compared to just two in 2017. The most accurately measured optical frequency ratio measurement ($q_{104}$) had a fractional uncertainty of $5.9\times 10^{-18}$~\cite{BACON2021}, with two others determined in the same campaign ($q_{102}$ and $q_{103}$) also reaching fractional uncertainties below $1\times 10^{-17}$.

\section{Correlation coefficients}
\label{sec:Correlations}

In contrast to previous least-squares adjustments performed by the WGFS, detailed consideration was given to correlations between the individual frequency ratio measurements, following the guidelines in~\cite{ROCIT_Guidelines}, to ensure that the frequency values obtained from the analysis were unbiased and that their uncertainties were not underestimated. In total, 483 correlation coefficients were calculated and included in the analysis. These fell into two categories: firstly correlations arising through the use of the same PFS or SFS to access the SI second and secondly others that were computed on an {\em ad-hoc} basis. These categories are considered in the following sections~\ref{sec:CorrPSFS} and~\ref{sec:CorrAdHoc} respectively.

\begin{figure}
\includegraphics[width=.9\columnwidth]{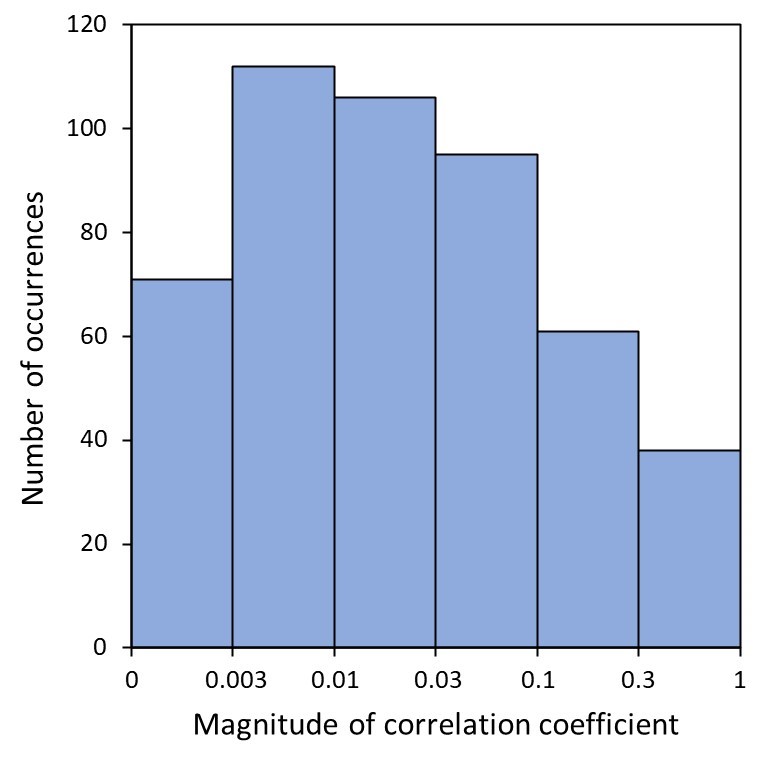}
\caption{\label{fig:CorrHist}Distribution of the magnitude of the 483 correlation coefficients included in the 2021 least-squares adjustment. Note the nonlinear scale used to display the magnitudes.}
\end{figure}

The distribution of the magnitude of the correlation coefficients is shown in figure~\ref{fig:CorrHist}. Of the 483 coefficients, 300 have a magnitude greater than 0.01, while 99 have a magnitude greater than 0.1.

\subsection{Correlation coefficients arising through access to the SI second}
\label{sec:CorrPSFS}

In this section, we estimate the correlation between absolute frequency measurements due to using the same PFS or SFS to access the SI second~\cite{ROCIT_Guidelines}. 
A procedure was developed to cover the measurements accessing the SI second through the duration of the TAI scale interval, $d_{\rm TAI}$, published monthly by the BIPM in Circular T, in which case correlation arises through the primary and secondary frequency standards (PSFS) common to the different $d_{\rm TAI}$ estimations. The same scheme can also determine the correlation between one measurement using $d_{\rm TAI}$ and one using a local PSFS (or an ensemble thereof) to access the SI second, provided that the local PSFS is operated in similar conditions when performing the local comparisons and when operating for TAI reports. It can similarly be used for the correlation between two measurements using the same local standards. However, in those cases it is likely that other, possibly more important, sources of correlation are present so that those cases were generally covered by a specific study (see section \ref{sec:CorrAdHoc}).

For practical reasons the procedure was applied starting in January 2014. This choice of start date ensured consistency in the set of primary standards participating in TAI and in the provision of information relative to $d_{\rm TAI}$. It also eased the task of obtaining necessary information from the operators of PSFS as well as from those performing the measurements. Furthermore, it met the need to cover the most accurate absolute frequency measurements. The most recent measurements considered were carried out in March 2020. 

Correlation between two measurements originating from using the same standards to access the SI second arises from the part of the standards' systematic uncertainty that is correlated from month to month, here denoted $u_{\rm bS}$, when the two measurements are performed in different months. When the two measurements occur during the same month, correlation is from the total uncertainty of the standards. The stationary part $u_{\rm bS}$ is generally not readily available and a specific study is necessary for each standard. Because the frequency standards at PTB and LNE-SYRTE typically contribute about 90$\,$\% of $d_{\rm TAI}$ over the period of interest and are also used as the local reference for a majority of the measurements considered here, their operators were approached for specific determinations of $u_{\rm bS}$ (table~\ref{tab:uBcor}). Note that, in the case of SYRTE-FORb, the recommended uncertainty of the secondary representation of the second $u_{\rm Srep}$ was added to the values in table~\ref{tab:uBcor} to compute $u_{\rm bS}$. For all other PSFS not included in table~\ref{tab:uBcor}, $u_{\rm bS}$ is taken to be $u_{\rm b}$.

\begin{table}
\caption{\label{tab:uBcor}Values of the stationary part $u_{\rm bS}$ of the systematic uncertainty of the PTB and SYRTE frequency standards over the years 2014 to 2020. For the PTB standards these were provided by S. Weyers, and for the SYRTE standards by M. Abgrall and L. Lorini.}
\begin{indented}
\item[]
\begin{tabular}{@{}lll}
\br
Standard      & Period        & $u_{\rm bS}/10^{-16}$  \\
\mr
PTB-CSF1      & Jan 2014 -- Feb 2019 & 3.1 \\
              & Mar 2019 -- Dec 2020 & 2.0 \\
\ms
PTB-CSF2      & Jan 2014 -- Dec 2020 & ${\rm Min}(2.1, u_{\rm b})$\\
\ms
SYRTE-FO1     & Jan 2014 -- Feb 2018 & 2.5 \\
              & Mar 2018 -- Aug 2018 & 2.3 \\
              & Sep 2018 -- Dec 2020 & 1.7 \\
\ms
SYRTE-FO2     & Jan 2014 -- Sep 2015 & 1.8 \\
              & Oct 2015 -- Feb 2018 & 1.5 \\
              & Mar 2018 -- Dec 2020 & 1.2 \\
\ms
SYRTE-FORb    & Jan 2014 -- Sep 2015 & 2.3 \\
              & Oct 2015 -- Feb 2018 & 2.1 \\
              & Mar 2018 -- Dec 2020 & 1.8 \\
\br
\end{tabular}
\end{indented}
\end{table}

The general equation to compute the correlation between measurements $q_x$ and $q_y$ due to using common PSFS in accessing the SI second is
\begin{equation} \label{eq:CoeffSI}
r(q_x, q_y)= \frac{\sum_{i_x,i_y} {w_{i_x} w_{i_y} \sum_k { w_{i_x,k}u_{{\rm bS}_{i_x,k}} w_{i_y,k}u_{{\rm bS}_{i_y,k}} }} } {u_x u_y}
\end{equation}
where $i_x$ and $i_y$ index the months used to access the SI second for measurements $q_x$ and $q_y$ and $w_{i_x}$ and $w_{i_y}$ are the weights of month $i_x$ and $i_y$ respectively. The PSFS are labelled with the index $k$. 

For measurements accessing the SI second through $d_{\rm TAI}$
$w_{i,k}$ is the weight of standard $k$ in the estimation of $d_{\rm TAI}$ for month $i$. 
The set of weights $w_{i,k}$ for all standards for each month $i$ over the period January 2014 to March 2020 (MJD 56659--58939) is collected from the monthly files etyy.mm (where yy.mm identifies the month), available on the BIPM ftp server~\cite{BIPM_EALWeb}, which provide the fractional frequency of the free atomic time scale EAL (\'Echelle Atomique Libre) as estimated from primary and secondary frequency standards.

For measurements accessing the SI second through local standards, $w_{i,k}$ is the weight of each local PSFS used in the comparisons.
In some cases, exact information about these weights was obtained from the groups that performed the measurements, otherwise they were estimated from the publication (e.g. a plot of residuals indicating dates and uncertainties).

The total uncertainties of the two measurements are $u_x$ and $u_y$. 
However, when $i_x = i_y = i$ (two measurements made in a single common month), correlation is through the total uncertainty $u_{i,k}$ of the standard $k$ in the estimation of $d_{\rm TAI}$ for this month, not through $u_{{\rm bS}_{i,k}}$. In all cases $u_{\rm bS}$ may vary with time so the month of operation needs to be specified.
 
This computation was carried out for 34 absolute frequency measurements, 33 of which took place in the period considered. Measurement $q_{46}$~\cite{LeTargat2013} was added even though it dates from 2010--2011 because of its small uncertainty and significant correlation with several more recent measurements, To compute correlation coefficients involving $q_{46}$, the $u_{\rm bS}$ values of January 2014 were used for the relevant SYRTE standards. Similarly some of the numerous measurements of $q_{56}$ were taken before 2014, and were assigned to the first months of 2014 by giving a larger weight to those months. The access to the SI second was modelled as follows (see table \ref{tab:Listmeas} for more details):

\begin{table}
\caption{\label{tab:Listmeas}List of absolute frequency measurements with relevant information used to compute correlation from access to the SI second.}
\begin{indented}
\item[]

\begin{tabular}{@{}lllll}
\br
      &            &      & No. of &  \\
Label & Transition & Lab.& months & Standards \\
\mr
$q_3$ & $\nu_1$ & NICT & 2 & TAI \\
$q_7$ & $\nu_3$ & SYRTE & 2 & FO2 \\
$q_{14}$ & $\nu_6$ & NPL & 2 & CsF2 \\
$q_{18}$ & $\nu_7$ & NPL & 2 & CsF2 \\
$q_{25}$ & $\nu_8$ & KRISS & 1 & TAI \\
$q_{32}$ & $\nu_{10}$ & NRC & 2 & TAI \\
$q_{34}$ & $\nu_{11}$ & Torun & 1 & TAI \\
$q_{35}$ & $\nu_{11}$ & Torun & 1 & TAI \\
$q_{43}$ & $\nu_{12}$ & NMIJ & 1 & TAI \\
$q_{44}$ & $\nu_{12}$ & NIM & 2 & NIM5 \\
$q_{46}$ & $\nu_{12}$ & SYRTE & 1$^\ast$ & FO1,FO2,FOM \\
$q_{47}$ & $\nu_{12}$ & SYRTE & 4 & FO1,FO2,FOM \\
$q_{48}$ & $\nu_{12}$ & PTB & 1 & CSF2 \\
$q_{49}$ & $\nu_{12}$ & PTB & 1 & CSF1,CSF2 \\
$q_{50}$ & $\nu_{12}$ & NICT & 3 & TAI \\
$q_{55}$ & $\nu_{13}$ & Wuhan & 4 & TAI \\
$q_{56}$ & $\nu_{14}$ & SYRTE & 37$^{\ast}$ & TAI \\
$q_{58}$ & $\nu_{14}$ & SYRTE & 1 & FO1,FO2,\\
         &            &       &   & PTB-CSF1, \\
         &            &       &   & PTB-CSF2 \\
$q_{70}$ & $\nu_8$ & ECNU & 1 & TAI \\
$q_{71}$ & $\nu_7$ & NPL & 1 & TAI \\
$q_{72}$ & $\nu_{12}$ & NPL & 1 & TAI \\
$q_{73}$ & $\nu_{12}$ & PTB & 10 & CSF1,CSF2 \\
$q_{74}$ & $\nu_1$ & NICT & 1 & TAI \\
$q_{75}$ & $\nu_8$ & NIST & 8 & TAI \\
$q_{76}$ & $\nu_8$ & INRIM & 5 & TAI \\
$q_{88}$ & $\nu_{13}$ & Wuhan & 1 & TAI \\
$q_{89}$ & $\nu_8$ & NMIJ & 6 & TAI \\
$q_{90}$ & $\nu_{12}$ & NICT & 14 & TAI$^{\ast}$ \\
$q_{91}$ & $\nu_{12}$ & INRIM & 2 & CsF2 \\
$q_{96}$ & $\nu_{12}$ & NIST & 4 & TAI \\
$q_{97}$ & $\nu_4$ & NIST & 5 & TAI \\
$q_{98}$ & $\nu_7$ & PTB & 10 & CSF1,CSF2 \\
$q_{105}$ & $\nu_{13}$ & NIM & 1 & NIM5 \\
\br
\end{tabular}
\item[]$^\ast$See text for further details.
\end{indented}
\end{table}

\begin{itemize}
\item For 19 measurements accessing the SI second through $d_{\rm TAI}$, by specifying the list of months along with the weight assigned to each month in the determination. Of these, nine measurements correspond to the simple case where the measurement was performed in a single month. 
\item For 14 measurements accessing the SI second through local standards, by assigning the ensemble of individual comparisons to a set of months and estimating for each month a weight which was then shared between the local PSFS used during that month.  
Note that measurement $q_{58}$ (an absolute frequency measurement of the SYRTE Rb fountain) included comparisons to the remote PTB Cs fountains in addition to the local Cs fountains.  
\item In the specific case of $q_{90}$~\cite{Nemitz2021}, the authors estimated 63 comparisons of their optical clock to 8 individual PFS using data published by the BIPM. Because the PFS used in the comparisons are those providing the estimation of $d_{\rm TAI}$ and their contributions to the absolute frequency measurement were adequately provided with monthly values in figure 4b of~\cite{Nemitz2021}, this measurement was introduced as if accessing through 14 monthly $d_{\rm TAI}$ values with the monthly weights taken from figure 4b. 
\end{itemize}

The computation of correlation coefficients also took into account revised total uncertainties for some absolute frequency measurements, as detailed in section \ref{sec:InputMods}. This concerns $q_{73}$ and $q_{98}$ following a specific computation (section~\ref{sec:Unphysical_correlations}), and $q_{74}$, $q_{88}$ and $q_{105}$ for which the uncertainty was enlarged (see section~\ref{sec:Improved_consistency}). The computation yielded 561 coefficients, 399 of which were larger than 0.001 and which were used in the least-squares analysis. Two of them ($r(q_{14}, q_{18})$ and $r(q_{50}, q_{90})$) were not used because a specific calculation provided a more accurate estimation (see~\ref{sec:Ybion_NPL_PTB} and \ref{sec:NICT_Inion_Sr}). The complete list of correlation coefficients may be accessed at~\cite{Data_Repository}.

\subsection{Correlation coefficients computed on an ad-hoc basis}
\label{sec:CorrAdHoc}

Based on a review of the input data, a total of 86 additional correlation coefficients were identified to be potentially significant and were therefore computed on an ad-hoc basis. These correlations originate from several different sources.

Firstly, significant correlations are likely to arise if an atomic clock participates in more than one frequency comparison during the same period. In this scenario, the correlation coefficient will typically have contributions coming from both the statistical and the systematic uncertainties of the common clock.
Measurements involving the same atomic clock, but performed at different times, may also be correlated if the systematic uncertainty of the clock is not re-evaluated, or is only partially re-evaluated, between the two measurements. Although the uncertainty budgets of optical clocks typically evolve more rapidly than in the case of the caesium fountain primary frequency standards considered in section~\ref{sec:CorrPSFS}, some instances of this type of correlation were identified. 
Secondly, correlations may arise between any clocks based on the same atomic transition, if the same theoretical or experimental values of atomic coefficients are used to correct for systematic frequency shifts such as Zeeman or blackbody radiation shifts. Finally, correlations associated with systematic corrections may also arise between measurements involving
clocks that are based on different atomic transitions, but that are located within the same laboratory. An example would be the case of the gravitational redshift correction, which would be largely common to any remote comparisons involving clocks in that laboratory. 

Further details about the computation of correlation coefficients arising from these sources are provided in~\ref{app:AdHocCorr}.

\section{Modifications to the input data}
\label{sec:InputMods}

Review of the input data by the WGFS resulted in one modification to an input frequency value, compared to the published value, and several modifications to the uncertainties of particular input values.

In evaluating the correlation coefficients arising through access to the SI second, it was realised that the absolute frequency measurement of $^{115}$In$^+$ reported in~\cite{Ohtsubo2017} was carried out in three sessions, but that only two of these used TAI as a reference, with the value obtained in the last session being based on the evaluation of the NICT $^{87}$Sr lattice clock ($q_{50}$) reported in~\cite{Hachisu2017a}. The group that performed the measurements was therefore asked by the WGFS to recompute the $^{115}$In$^+$ frequency using data only referenced to TAI, and the resulting value of 1$\,$267$\,$402$\,$452$\,$901$\,$049.8(7.5)$\,$Hz, rather than the published value listed in table~\ref{tab:2017_Absolutes}, was used for $q_3$ in the least-squares adjustment.

Modifications were also made to the uncertainties of a few measurements in the input data set. As already stated in section~\ref{sec:NewData}, the uncertainty of $q_{51}$ was multiplied by $10^6$ to effectively exclude it from the least-squares adjustment, because the same data was used to contribute to $q_{90}$. However modifications to uncertainties were also made for several other reasons:
\begin{enumerate}
\item To avoid otherwise unphysical values of correlation coefficients.
\item To handle particular data points that are identified as outliers, making them statistically more consistent with the overall body of data.
\item To take account of the sparsity of the input data for some atomic transitions included in the adjustment.
\end{enumerate}
These adjustments are described in more detail in sections~\ref{sec:Unphysical_correlations}--\ref{sec:Increased_uncertainty} below, and summarized in table~\ref{tab:Uncertainty_changes}.

\begin{table*}
\caption{\label{tab:Uncertainty_changes}Changes made to the uncertainties of the input data to the least-squares adjustment, compared to published uncertainties. Note that these differ in some cases from the changes made in 2017. 
The rationales listed are described in more detail in sections~\ref{sec:NewData} (exclude from fit),~\ref{sec:Unphysical_correlations} (unphysical correlation coefficients)),~\ref{sec:Improved_consistency} (improved consistency) and~\ref{sec:Increased_uncertainty} (sparsity of data).}
\begin{indented}
\item[]\begin{tabular}{@{}lll}
\br
Measurement label  & Change to published uncertainty  & Rationale\\ \\
\mr
$q_1$     & Multiplied by 3                    & Improved consistency \\
$q_{31}$  & Multiplied by 1.5                  & Sparsity of data  \\
$q_{51}$  & Multiplied by $10^6$                  & Exclude from fit  \\
$q_{52}$  & Multiplied by 6                    & Improved consistency \\
$q_{73}$  & Increased to $1.65\times 10^{-16}$ & Unphysical correlation coefficient \\
$q_{74}$  & Multiplied by 3                    & Sparsity of data \\
$q_{78}$  & Multiplied by 3                    & Sparsity of data \\
$q_{88}$  & Multiplied by 2                    & Sparsity of data \\
$q_{98}$  & Increased to $1.6\times 10^{-16}$  & Unphysical correlation coefficient \\
$q_{105}$ & Multiplied by 2                    & Sparsity of data  \\
\br
\end{tabular}
\end{indented}
\end{table*}

\subsection{Avoiding unphysical correlation coefficients}
\label{sec:Unphysical_correlations}

Unphysical correlation coefficients may arise if different uncertainties are used and reported for the same frequency standard during the same period, when that standard is used for different purposes.
Initial runs to determine correlation coefficients as described in section \ref{sec:CorrPSFS} revealed several large values, with a few of them larger than 1. These large coefficients mostly involved measurements $q_{73}$ and $q_{98}$, with the largest value for $r(q_{73}, q_{98})$. Both measurements correspond to the determination of an optical frequency with respect to PTB primary standards CSF1 and CSF2 from a few tens of comparisons carried out over the period 2017--2019. In both cases, the details of the comparisons are well documented, $q_{73}$ through Table II in \cite{Schwarz2020} and $q_{98}$ through specific files transmitted by the PTB group to the WGFS~\cite{Data_Repository}. 
It was realized that the systematic uncertainty $u_{\rm b,Cs}$ associated with PTB-CSF1 is in both cases significantly lower than the systematic uncertainty value $u_{\rm b}$ in the reports of the monthly evaluation of TAI frequency by PTB-CSF1~\cite{BIPM_PSFSReports} and also significantly lower than the stationary part of its systematic uncertainty, $u_{\rm bS}$, as defined in section \ref{sec:CorrPSFS}. On the other hand, the treatment for CSF2 is quite in line with the values of $u_{\rm b}$ reported for TAI and with the values of $\rm u_{bS}$ that we used to compute correlations.

Because of the lower $u_{\rm b,Cs}$ values for CSF1, it is not possible to use the standard equation~\ref{eq:CoeffSI} to compute correlations with $q_{73}$ and $q_{98}$. However, these are the two most accurate determinations of absolute frequencies to date and need to be fully taken into account. We therefore resolved to increase {\it a minima} the total uncertainty of both measurements to the values that would be obtained if replacing for PTB-CSF1 the $u_{\rm b,Cs}$ values by the $u_{\rm bS}$ values as defined in section \ref{sec:CorrPSFS}. The resulting total uncertainties, used in the final analysis, are $1.65\times10^{-16}$ for $q_{73}$ and $1.6\times10^{-16}$ for $q_{98}$, rather than the published uncertainties of $1.5\times10^{-16}$ and $1.3\times10^{-16}$ respectively. 

Note that the main effect of this change is a higher uncertainty for the CSF1 measurements, especially those taken before February 2019 when $u_{\rm bS} = 3.1\times10^{-16}$. It also  results in a lower weight for the CSF1 measurements in the determination of the absolute value of the optical frequencies, meaning that in principle the values could change. This change is estimated to be lower than $3\times10^{-17}$ for both measurements and it was therefore decided not to change the absolute frequency values for the 2021 adjustment. Using the modified total uncertainties, all correlation coefficients computed have sensible values, with the largest coefficient $r(q_{73}, q_{98})$ being 0.729.

\subsection{Improving consistency of the input data set}
\label{sec:Improved_consistency}

The input data set includes several measurements of some quantities, the consistency of which can be checked directly~\cite{Riehle2018}. However a more complete and rigorous evaluation of the consistency of the input data was performed by carrying out a preliminary least-squares adjustment to identify outliers in the input data set. 
This was achieved through inspection of the normalised residuals $\rho_i = (q_i - \hat{q}_i)/u_i$, where the $\hat{q}_i$ are the optimised frequency values obtained from the fit.

This analysis indicated that the biggest outlier in the complete input data set was $q_{52}$, with $\rho_{52}=-6.90$. To make $q_{52}$ statistically more consistent with other measurements, its published uncertainty was therefore multiplied by six in the input data set. 

The other significant ($>3\sigma$) outlier identified in this way was $q_1$, with $\rho_1 = -4.87$. For consistency with the previous 2017 adjustment, the published uncertainty of $q_1$ was therefore multiplied by three in the final input data file.

\subsection{Accounting for sparsity of the input data}
\label{sec:Increased_uncertainty}


The WGFS also request and take into account information pertaining to comparison of frequency standards based on the same atomic transition, but such data is not always available. 
As discussed by Riehle {\it et al.}~\cite{Riehle2018}, the WGFS have developed procedures which take a cautious approach to uncertainty estimation in the case that a particular recommended frequency value is determined by very few, or even a single, input frequency value. Historically, whilst the recommended frequency values were derived from only absolute frequency measurements, this involved applying enlargement factors to the published uncertainty (in the case of a single input value) or to the uncertainty of the weighted mean (in the case of two independent input values). Although these procedures were previously amended to some degree to take account of the availability of high accuracy direct frequency ratio measurements, making {\it ad-hoc} adjustments to the uncertainties of selected optimised frequencies from the least-squares adjustment is unsatisfactory because it leads to inconsistencies in the complete output data set, which includes frequency ratios between every pair of atomic transitions. For this reason we departed from prior practice and made selected adjustments to the input data instead, followed later by a global enlargement (by a factor of two) to the uncertainties of every ratio in the output data set~(discussed in more detail in section~\ref{sec:Results}). This change in procedure ensured that the output data set we obtain is internally self-consistent. 

The adjustments made to account for sparsity of the input data are as follows:
\begin{enumerate}
\item In the case of the $^{88}$Sr$^+$ optical clock, the only published data consisted of absolute frequency measurements, meaning that this transition is decoupled from the rest of the analysis. None of the available measurements were new since 2017. In this case the frequency obtained from the least-squares adjustment is equivalent to taking a weighted mean of the measurements, and is dominated by $q_{31}$, which has a fractional uncertainty of $5.4\times 10^{-16}$, compared to the next most precise measurement $q_{32}$ with fractional uncertainty $1.5\times 10^{-15}$. For this reason, in 2017, the uncertainty of the value from the fit was increased by a factor of three. To achieve a similar result in the 2021 update to the list of recommended frequency values, taking into account the global uncertainty enlargement by a factor of two (discussed in section~\ref{sec:Results}), we increased the uncertainty on the input data point $q_{31}$ by a factor of 1.5.
\item The $^{115}$In$^+$ clock transition frequency obtained from the least-squares adjustment is determined almost entirely by two defining measurements $q_{74}$ and $q_{78}$ with fractional uncertainties far lower than the other measurements involving this standard ($q_1$, $q_2$ and $q_3$). These two defining measurements, an absolute frequency measurement and an optical frequency ratio measurement against $^{87}$Sr, were performed during the same period in a single laboratory (NICT), and are strongly correlated, with a correlation coefficient $r(q_{74},q_{78}) = 0.859$. For this reason the published uncertainties of both $q_{74}$ and $q_{78}$ were multiplied by three in the final input data file, whilst keeping $r(q_{74},q_{78})$ unchanged.
\item For the optical clock transition in $^{40}$Ca$^+$, the frequency obtained from the final least-squares adjustment is dominated by two new absolute frequency measurements $q_{88}$ and $q_{105}$ originating from a single laboratory and from the same measurement period. The published uncertainties of these measurements were therefore multiplied by two in the input data set. A factor of two rather than three was used in this case, because published comparisons between two $^{40}$Ca$^+$ optical clocks~\cite{Huang2016, Huang2020} support the uncertainty evaluation for the clock used in these measurements. 
\end{enumerate}

No new data was available for either the 1S--2S transition in $^1$H or the optical clock transition in $^{40}$Ca. Since these two transitions are linked to the other input data only via Cs, their recommended frequencies and uncertainties should remain the same as in 2017. For convenience, in these two cases, we left the input data unchanged and simply removed the transitions from the output covariance matrix, after verifying that the least-squares adjustment did indeed give the same values for the transition frequencies as in 2017. 

Although no other adjustments were made to the published uncertainties in the input data, we note the following points:
\begin{itemize}
    \item The optimized frequency for the $^{199}$Hg optical clock transition is mainly determined by frequency ratio measurements from RIKEN involving other better-known optical frequencies, especially that of the $^{87}$Sr clock transition. However measurements from a second laboratory (LNE-SYRTE) were also available, and all normalised residuals were observed to have a magnitude less than one.
    \item In the case of the $^{27}$Al$^+$ optical clock transition, all measurements originate from a single laboratory (NIST) and the frequency obtained from the least-squares adjustment is mainly determined by measurements $q_{103}$ and $q_{104}$ which were made during the same measurement campaign and have a correlation coefficient $r(q_{103},q_{104})=0.329$. However all normalised residuals except for that for $q_8$, which has little weight in the adjustment, were observed to have a magnitude less than one.
    \item The optimized frequency for the $^{199}$Hg$^+$ optical clock transition is determined almost entirely by a single optical frequency ratio measurement $q_{62}$. The only absolute frequency measurement available, $q_9$, has a normalised residual with magnitude $>2$ but has little weight in the least-squares adjustment. 
    \item The optimized frequency for the E2 optical clock transition in $^{171}$Yb$^+$ is determined almost entirely by a frequency ratio measurement against the E3 transition, $q_{99}$, which has an uncertainty an order of magnitude lower than any of the other measurements involving the E2 transition. 
    \item Although three absolute frequency measurements of the optical clock transition in $^{88}$Sr were available, the frequency obtained from the least-squares adjustment is essentially completely determined by two measurements of the ratio between the $^{88}$Sr and $^{87}$Sr clock transitions ($q_{67}$ and $q_{77}$), which have similar fractional uncertainties ($2.3\times 10^{-17}$ and $2.8\times 10^{-17}$, respectively). These measurements were made in different groups, and differ by 1.5 times their combined relative uncertainties.
\end{itemize}

The focus in this analysis was on sparsity of input data that affects the recommended frequency values. Some optical frequency ratios are also determined by one or two input measurements, but no adjustments were made to input data in such cases, meaning that the uncertainties assigned to optical frequency ratio values may perhaps be less conservative than those assigned to the recommended frequency values.

\section{Results from the least-squares adjustment} 
\label{sec:Results}

Once the input data had been finalised, the analysis software was run for a last time to calculate optimised values for each frequency ratio and absolute frequency value, and to check the self-consistency of the input data. 
The results obtained using the different algorithms and software were compared to verify the results.
The results obtained using the two different implementations of the least-squares algorithms yielded absolute frequency values and frequency ratio values that differed at most by one in the least-significant (24th) digit of the computation, while the uncertainties were identical to the 4 significant figures computed. Slightly larger differences were seen in some cases between the results obtained using the two different algorithms. However even in this case the maximum differences observed in absolute frequency values or frequency ratio values were 2 parts in $10^{21}$ while uncertainties differed by no more than 2 in the least significant digit of the four computed. The output correlation coefficients between the absolute frequency values computed using the two algorithms agreed to better than 1 part in $10^5$. This level of agreement between the different algorithms and software is several orders of magnitudes below the uncertainties on the output values. When appropriate truncation and rounding is applied to obtain recommended values and uncertainties, all computations give identical recommended values and uncertainties.

\begin{figure}
    \centering
    \includegraphics[width=\columnwidth]{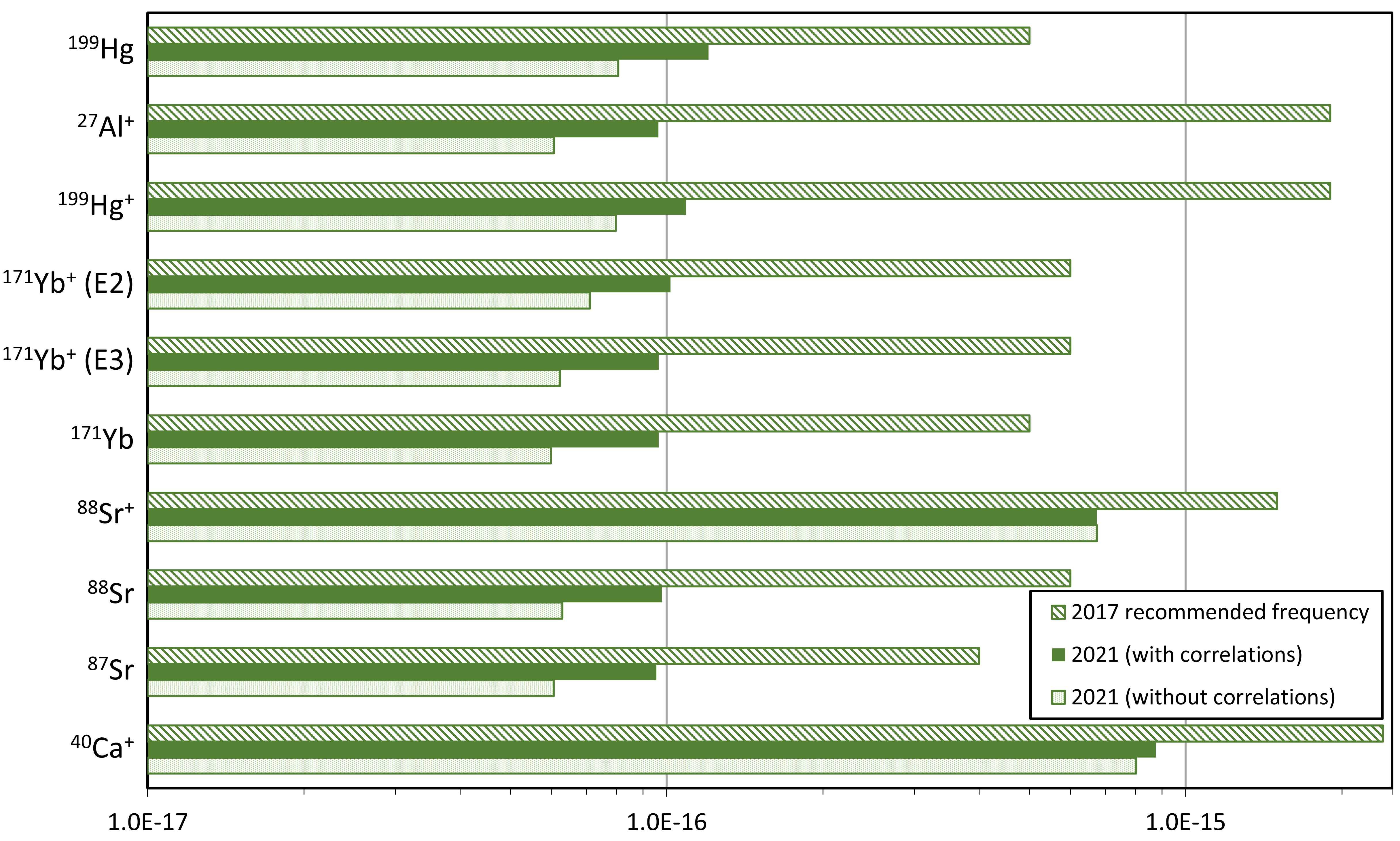}
    \caption{\label{fig:EffectOfCorrelations}Uncertainty of the 2021 adjusted frequency values for ten optical frequency standards, plotted on a logarithmic scale, showing that the effect of neglecting correlations is to underestimate the uncertainty. Also shown for comparison are the uncertainties of the 2017 recommended frequency values, though it should be noted that these are typically larger than the uncertainty from the 2017 fit, due to the cautious approach to uncertainty estimation taken by the WGFS.}
\end{figure}

\begin{figure*}
    \centering
    \includegraphics[width=1.8\columnwidth]{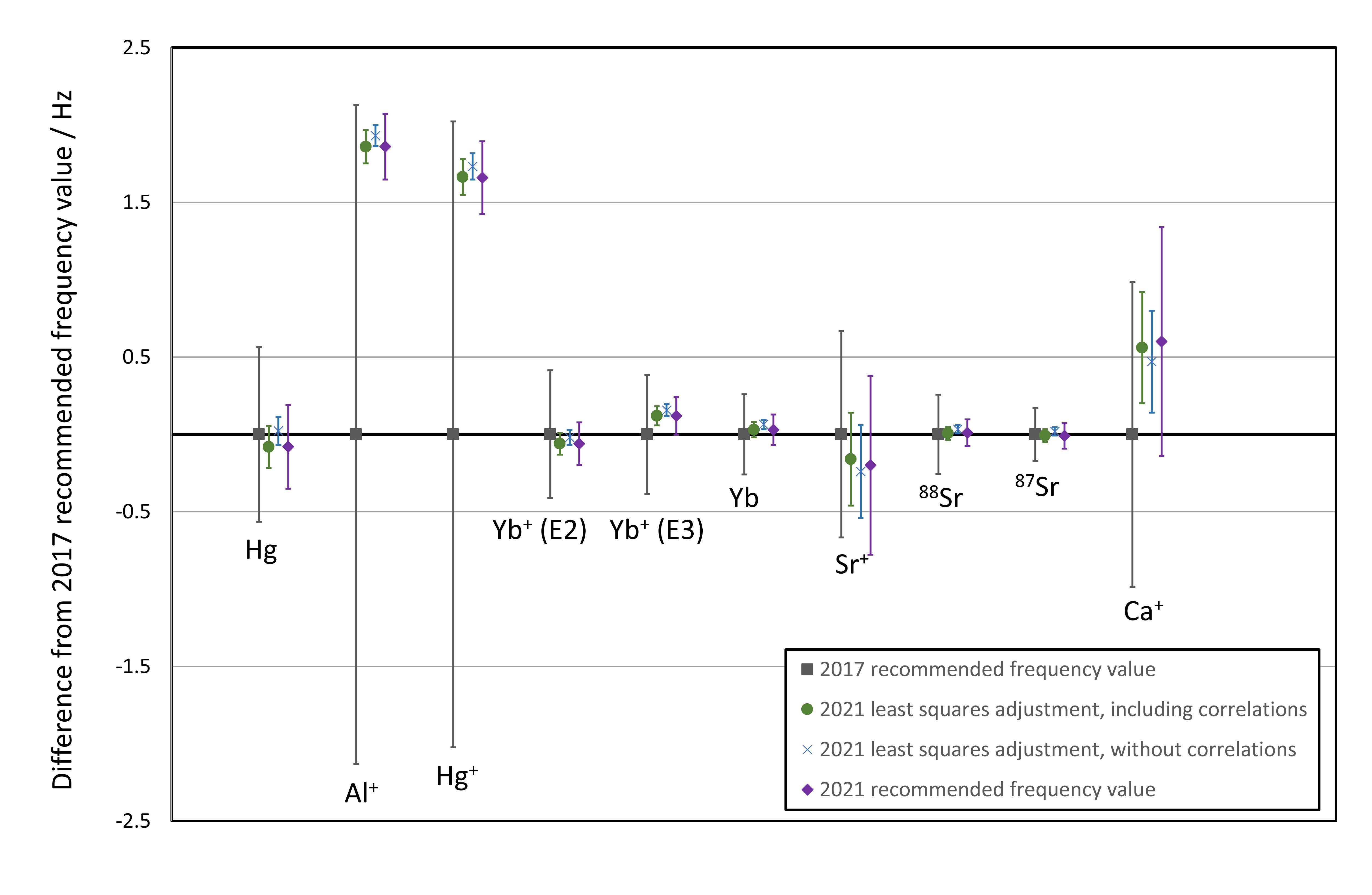}
    \caption{\label{fig:RecommendedValues}2021 adjusted frequency values and uncertainties for ten optical frequency standards (green circles), showing that neglecting correlations (blue crosses) can bias the frequency obtained from the fit. Also shown are both the 2017 and 2021 recommended frequency values and uncertainties (grey squares and purple diamonds, respectively). All ten optical frequency standards are now secondary representations of the second. Note that small differences observed between some of the 2021 recommended frequency values and the values obtained from the 2021 least-squares adjustment are due to rounding.}
\end{figure*}

The importance of accounting for correlations in the analysis is clearly visible in figures~\ref{fig:EffectOfCorrelations} and~\ref{fig:RecommendedValues}: including correlations leads to uncertainties up to 60\% higher than if correlations are neglected and shifts the adjusted frequency values by up to 70\% of their uncertainty. The Birge ratio for the final least-squares adjustment (including correlations) was 1.064  and the goodness-of-fit was estimated to be 0.18, providing an important check on the overall consistency of the set of input measurements (with the modifications described in section~\ref{sec:InputMods}).

Bearing in mind the purpose and applications of the list of recommended frequency values, in particular the use of secondary representations of the second for calibration of the scale interval of TAI, the WGFS is concerned to avoid any discontinuities in the values, i.e. to ensure that any changes between adjustments have a magnitude less than the combined uncertainties of the old and new values. This motivates a cautious approach to the derivation of uncertainties, based on the informed judgement of the membership, as previously elucidated in [1].

The least-squares adjustment procedure used to analyse the over-constrained input data set implicitly assumes that the probability density of the input data is well approximated by a Gaussian distribution, which is not well justified or easily tested in cases where the output values are determined by just one or two input measurements. Furthermore, it assumes that correlations between the input measurements are precisely known and taken into account. Although significant effort was devoted to identifying and estimating such correlation coefficients in the 2021 analysis, it remains the case that as-yet-unidentified sources of uncertainty or correlations may exist.

For this reason the WGFS maintained the previous cautious approach to uncertainty estimation, but departed from previous practice by applying a global expansion factor to the output covariance matrix rather than applying separate expansion factors to the uncertainties of selected recommended frequency values. This global expansion factor achieves the important outcome that the set of recommended frequency values and the ratios between them are internally self-consistent. On this occasion the expansion factor selected corresponded to a multiplication by two of the raw fit uncertainties. The recommended frequency value itself is the result obtained directly from the least-squares adjustment, rounded as appropriate given the magnitude of the recommended uncertainty. The uncertainties of the recommended frequency values should be understood and used as estimated standard uncertainties, corresponding to a coverage interval of 68.27$\,$\%, i.e. the expansion factor applied is intended to allow for as-yet-unidentified sources of uncertainty or correlation, and is separate from any coverage factor used to calculate expanded uncertainties or confidence intervals as described in~\cite{BIPM_GUM}.

The global expansion factor by a factor of two was also noted by the WGFS to have the effect, considered desirable, of avoiding recommended frequency values with uncertainties significantly lower than the uncertainty of any individual realisation of the SI second so far.

\begin{table*}
\caption{\label{tab:2021RecValues}The twelve recommended frequency values updated in March 2021~\cite{CCTF_Rec2_2021}.}
\begin{indented}
\item[]\begin{tabular}{@{}llll}
\br
Transition label & Atomic species & 2021 recommended frequency value / Hz & Recommended fractional uncertainty \\
\mr
$\nu_1$  & $^{115}$In$^+$      & 1267\,402\,452\,901\,041.3                       & $4.3\times 10^{-15}$ \\
$\nu_3$  & $^{199}$Hg          & 1128\,575\,290\,808\,154.32                      & $2.4\times 10^{-16}$ \\
$\nu_4$  & $^{27}$Al$^+$       & 1121\,015\,393\,207\,859.16                      & $1.9\times 10^{-16}$ \\
$\nu_5$  & $^{199}$Hg$^+$      & 1064\,721\,609\,899\,146.96                      & $2.2\times 10^{-16}$ \\
$\nu_6$  & $^{171}$Yb$^+$ (E2) & {\lineup\0}688\,358\,979\,309\,308.24            & $2.0\times 10^{-16}$ \\
$\nu_7$  & $^{171}$Yb$^+$ (E3) & {\lineup\0}642\,121\,496\,772\,645.12            & $1.9\times 10^{-16}$ \\
$\nu_8$  & $^{171}$Yb          & {\lineup\0}518\,295\,836\,590\,863.63            & $1.9\times 10^{-16}$ \\
$\nu_{10}$ & $^{88}$Sr$^+$     & {\lineup\0}444\,779\,044\,095\,486.3             & $1.3\times 10^{-15}$ \\
$\nu_{11}$ & $^{88}$Sr         & {\lineup\0}429\,228\,066\,418\,007.01            & $2.0\times 10^{-16}$ \\
$\nu_{12}$ & $^{87}$Sr         & {\lineup\0}429\,228\,004\,229\,872.99            & $1.9\times 10^{-16}$ \\
$\nu_{13}$ & $^{40}$Ca$^+$     & {\lineup\0}411\,042\,129\,776\,400.4             & $1.8\times 10^{-15}$ \\
$\nu_{14}$ & $^{87}$Rb         & {\lineup\0\0\0\0\0\0\;}6834\,682\,610.904\,3126  & $3.4\times 10^{-16}$ \\
\br
\end{tabular}
\end{indented}
\end{table*}

The updated recommended frequency values and uncertainties were approved at the 22nd meeting of the CCTF in March 2021~\cite{CCTF_Rec2_2021}, and are listed in table~\ref{tab:2021RecValues}. At the same time, two additional reference transitions (in $^{88}$Sr and $^{40}$Ca$^+$) were approved as optical secondary representations of the second, bringing the total to ten. These additions reflected the improved uncertainties for their recommended frequency values and the fact that prospects for using standards based on these atomic transitions for future contributions to TAI were considered to be good.

For the ten optical secondary representations of the second, the 2021 recommended frequency values are compared with the 2017 values in figure~\ref{fig:RecommendedValues}. The most significant changes in the recommended frequency values are for $^{27}$Al$^+$ and $^{199}$Hg$^+$, and highlight the benefit of the cautious approach to uncertainty assignment traditionally taken by the WGFS. It is worth noting that the uncertainty of the recommended frequency value for $^{199}$Hg$^+$ is significantly reduced even though there is no new data for this transition. This is because in 2017, the $^{27}$Al$^+$/$^{199}$Hg$^+$ ratio essentially determined the absolute frequency of $^{27}$Al$^+$ and had almost no effect on the absolute frequency of $^{199}$Hg$^+$, while the new data available for $^{27}$Al$^+$ in 2021 means that the $^{27}$Al$^+$/$^{199}$Hg$^+$ ratio now reduces the uncertainty of the $^{199}$Hg$^+$ frequency.

\begin{figure}
    \centering
    \includegraphics[width=\columnwidth]{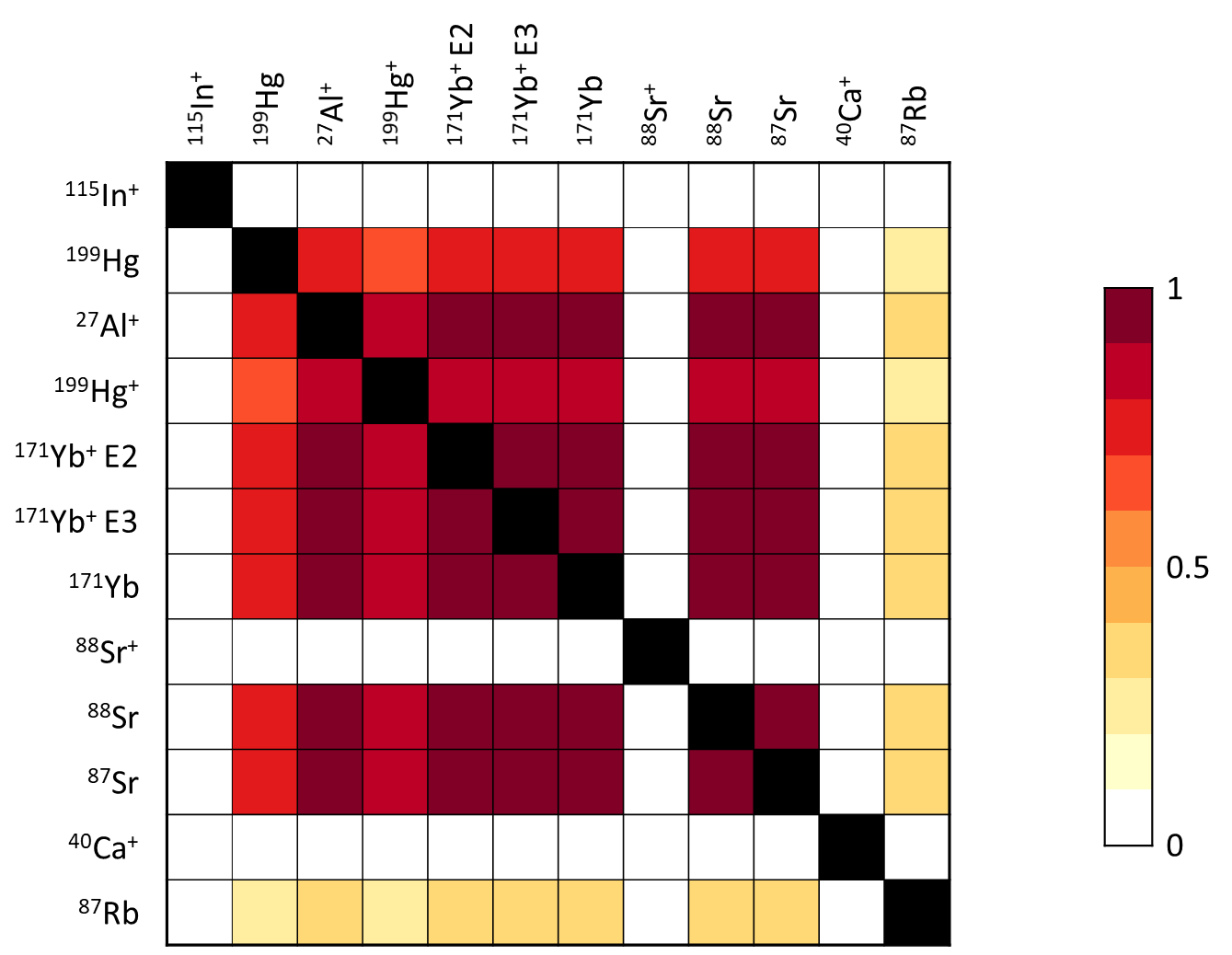}
    \caption{\label{fig:OutCorrHeatMap}Visualisation of the correlation matrix between the 2021 recommended frequency values. The colour in the heat map indicates the size of the correlation coefficient between each pair of recommended frequency values.}
\end{figure}

The effect of the increased number of optical frequency ratio measurements in the input data set, many of which have uncertainties significantly smaller than any measurement involving caesium primary frequency standards, is to create strong correlations between most of the recommended frequency values, as indicated in figure~\ref{fig:OutCorrHeatMap}. With the exception of correlation coefficients involving $^{88}$Sr$^+$ and $^{40}$Ca$^+$, all correlation coefficients between recommended frequency values for optical secondary representations of the second are greater than 0.65, and ten are greater than 0.95. 
To calculate any other frequency ratio from the recommended frequency values, it is necessary to take account of these correlations by using the output covariance matrix in order to compute the uncertainty of that frequency ratio correctly~\cite{BIPM_GUM}.
For completeness, we therefore list values and uncertainties for each frequency ratio in~\ref{sec:OutputRatios}.
The need to take correlations into account similarly applies to any other quantities depending on or combining these frequency ratios.

\section{Conclusion}
\label{sec:Conclusion}

Including correlations in the analysis underpinning the 2021 update to the list of recommended frequency values, whilst essential for a proper treatment of the data, significantly increased the effort required from the group performing the analysis on behalf of the WGFS. This is because the vast majority of the correlation coefficients were estimated by this group. In many cases the publications describing the measurements do not contain sufficient detail to identify all potential sources of correlation, and although efforts were made to gather this information when measurements were submitted for consideration by the WGFS, time constraints limited the amount of follow-up interaction that was possible. To prepare for future updates to the list, it would be beneficial to invest time in improving practices that will facilitate the computation~\cite{CCTF_Rec1_2021}, for example by encouraging groups to compute and submit correlation coefficients themselves, at least between different measurements performed within any given institution, or within coordinated comparison campaigns. 

As a result of the 2021 update to the list of recommended frequency values, six optical standards now have recommended uncertainties of $1.9$--$2.0\times 10^{-16}$, essentially the same as that of the best caesium fountain primary frequency standards. This establishes a solid link between optical frequency standards and the current definition of the second, which will be important to ensure continuity at the point of a future redefinition.
Prior to that redefinition, it also increases the weight that optical secondary representations of the second can have in the steering of TAI. 
In recent years, the very best reported evaluations of the frequency of TAI with SFS have a total uncertainty (excluding $u_{\rm Srep}$) of order $2\times 10^{-16}$. With the uncertainties from the 2017 adjustment their weight in estimating $d_{\rm TAI}$ was typically about 4$\,$\% for $^{171}$Yb clocks and about 6$\,$\% for $^{87}$Sr clocks resulting, in the case of three evaluations in a month, in a typical total contribution to $d_{\rm TAI}$ of about 10$\,$\%. The uncertainties from the 2021 adjustment were first used in April 2022 and the typical weight of individual contributions is now between 10$\,$\% and 16$\,$\% with the total weight for three evaluations per month being between 16$\,$\% and 26$\,$\%. 
The number of evaluations is also tending to increase so that SFS may eventually contribute as much as Cs fountains in estimating $d_{\rm TAI}$. Note that, in a future adjustment of the standard frequency values, this large contribution of SFS to $d_{\rm TAI}$ will complicate the evaluation of correlations for the cases when $d_{\rm TAI}$ is used to access the SI second. In the present work we could neglect this effect as the contributions of SFS were quite rare before March 2020 (with the notable exception of the SYRTE $^{87}$Rb fountain) and they very rarely contributed more than 10$\,$\% in any given month.

Optical frequency ratio measurements have a vital role to play in verifying the international consistency of optical clocks at a level better than $5\times 10^{-18}$, one of the criteria set out in the international roadmap towards the redefinition of the SI second~\cite{Dimarcq2023}. 
In the 2021 update to the recommended frequency values, the rules and criteria previously employed by the WGFS were modified to ensure that the output data from the least-squares adjustment includes a complete set of frequency ratio values (including absolute frequencies as a special case) that are internally self-consistent. 
Two different options are currently envisaged for a redefinition of the second, which might be based on a single reference optical frequency in a similar way to today's caesium-based definition, or might instead be based on an ensemble of reference frequencies as proposed in~\cite{Lodewyck2019}. Whichever option is eventually selected, one or more of the frequency ratios from a similar least-squares adjustment will be used to set the defining constant or constants appearing in the new definition, and the frequency ratios will also play an important role in the {\it Mise en pratique} for the new definition. Increasing scrutiny will therefore need to be paid to the evolution of these frequency ratio values in subsequent updates to the list of recommended frequency values, to ensure the stability of the new definition and realisation of the second.

\ack   

We thank S. Weyers, M. Abgrall, L. Lorini, M. Pizzocaro, D. B. Hume, N. Nemitz, H. Hachisu, N. Ohtsubo and T. Kobayashi for providing information to support the evaluation of correlation coefficients, as well as P. Tavella, N. Dimarcq and members of the CCL-CCTF WGFS for helpful discussions. 
HM and SB acknowledge funding from the European Metrology Programme for Innovation and Research (EMPIR) project 18SIB05 ROCIT, co-financed by the Participating States and from the European Union's Horizon 2020 research and innovation programme. HM also acknowledges funding from the UK Department for Science, Innovation and Technology as part of the National Measurement System Programme, and thanks the BIPM for supporting a one-month secondment.

\appendix   



\section{Details of correlation coefficients computed on an ad-hoc basis}
\label{app:AdHocCorr}

In this appendix, we provide details of the 86 correlation coefficients that were computed on an {\it ad-hoc} basis. The values of these correlation coefficients are listed in table~\ref{tab:AdHocCorrelations}.

\begin{table*}
\caption{\label{tab:AdHocCorrelations}Correlation coefficients computed on an ad-hoc basis.}
\begin{indented}
\item[]\begin{tabular}{@{}lllll}
\br
\multicolumn{5}{l}{Measurements involving $^{171}$Yb$^+$ and $^{87}$Sr optical clocks at PTB} \\
\mr
$r(q_{19},q_{45})=0.981$ & $r(q_{73},q_{92})=-0.060$ & $r(q_{92},q_{98})=0.002$ & $r(q_{92},q_{99})=-0.009$ & $r(q_{98},q_{99})=-0.002$ \\
\mr
\multicolumn{5}{l}{Measurements involving $^{171}$Yb$^+$ optical clocks at NPL and PTB} \\
\mr
$r(q_{13},q_{14})=0.030$ & $r(q_{13},q_{63})=\lineup\m 0.060$ & $r(q_{14},q_{18})=0.680$ & $r(q_{14},q_{63})=\lineup\m 0.507$ & $r(q_{16},q_{18})=\lineup\m 0.004$ \\
$r(q_{16},q_{19})=0.006$ & $r(q_{16},q_{63})=-0.007$ & $r(q_{18},q_{19})=0.009$ & $r(q_{18},q_{63})=-0.018$ & $r(q_{19},q_{63})=-0.015$ \\
\mr
\multicolumn{5}{l}{Measurements performed in June 2015} \\
\mr
$r(q_{7},q_{60})\;\;\;=\lineup\m 0.449$ & $r(q_{7},q_{61})\;\;\;=\lineup\m0.221$ & $r(q_{47},q_{60})\;\;\;=-0.033$ & $r(q_{47},q_{69})=\lineup\m0.018$ & $r(q_{47},q_{84})\;\,=-0.018$ \\
$r(q_{47},q_{85})\;\,=-0.022$ & $r(q_{47},q_{100})=-0.026$ & $r(q_{47},q_{101})\;\,=-0.022$ & $r(q_{49},q_{86})=-0.004$ & $r(q_{49},q_{87})\;\,=-0.007$ \\
$r(q_{58},q_{61})\;\,=-0.636$ & $r(q_{58},q_{69})\;\,=-0.713$ & $r(q_{60},q_{61})\;\;\;=\lineup\m0.456$ & $r(q_{60},q_{69})=-0.028$ & $r(q_{60},q_{84})\;\,=\lineup\m 0.028$ \\
$r(q_{60},q_{85})\;\,=\lineup\m 0.035$ & $r(q_{60},q_{100})=\lineup\m0.041$ & $r(q_{60},q_{101})\;\,=\lineup\m0.035$ & $r(q_{61},q_{69})=\lineup\m0.597$ & $r(q_{69},q_{84})\;\,=-0.015$ \\
$r(q_{69},q_{85})\;\,=-0.019$ & $r(q_{69},q_{100})=-0.022$ & $r(q_{69},q_{101})\;\,=-0.019$ & $r(q_{71},q_{84})=\lineup\m0.092$ & $r(q_{71},q_{85})\;\,=\lineup\m0.112$ \\
$r(q_{71},q_{86})\;\,=\lineup\m 0.086$ & $r(q_{71},q_{87}))=\lineup\m0.151$ & $r(q_{84},q_{85})\;\;\;=\lineup\m0.155$ & $r(q_{84},q_{86})=\lineup\m0.105$ & $r(q_{84},q_{87})\;\,=\lineup\m0.183$ \\
$r(q_{84},q_{100})=\lineup\m 0.117$ & $r(q_{84},q_{101})=\lineup\m0.100$ & $r(q_{85},q_{86})\;\;\;=\lineup\m0.128$ & $r(q_{85},q_{87})=\lineup\m0.224$ & $r(q_{85},q_{100})=\lineup\m0.027$ \\
$r(q_{85},q_{101})=\lineup\m 0.023$ & $r(q_{86},q_{87}))=\lineup\m0.178$ & $r(q_{100},q_{101})=\lineup\m0.027$ &  &  \\
\mr
\multicolumn{5}{l}{Optical frequency ratio measurements at NIST/JILA$^a$} \\
\mr
$r(q_{102},q_{103})=0.615$ & $r(q_{102},q_{104})=-0.207$ & $r(q_{103},q_{104})=0.329$ & & \\
\mr
\multicolumn{5}{l}{Measurements involving the INRIM $^{171}$Yb lattice clock$^b$} \\
\mr
$r(q_{24},q_{81})=0.088$ & $r(q_{76},q_{82})=0.191$ & $r(q_{76},q_{83})=0.175$ & $r(q_{82},q_{83})=0386$ & \\
\mr
\multicolumn{5}{l}{Measurements involving optical clocks at RIKEN$^c$} \\
\mr
$r(q_{59},q_{79})=0.826$ & $r(q_{66},q_{79})=-0.009$ &  &  &  \\
\mr
\multicolumn{5}{l}{Measurements involving the $^{115}$In$^+$ and $^{87}$Sr optical clocks at NICT} \\
\mr
$r(q_{3},q_{74})\;\,=0.026$ & $r(q_{3},q_{78})\;\,=\lineup\m0.028$ & $r(q_{41},q_{90})=\lineup\m0.001$ & $r(q_{50},q_{90})=0.061$ & $r(q_{74},q_{78})=0.859$ \\
$r(q_{78},q_{80})=0.006$ & $r(q_{78},q_{90})=-0.021$ & $r(q_{80},q_{90})=-0.026$ &  &  \\
\mr
\multicolumn{5}{l}{Measurements involving the $^{171}$Yb and $^{87}$Sr optical lattice clocks at NMIJ$^d$} \\
\mr
$r(q_{78},q_{94})=0.005$ & $r(q_{89},q_{93})=0.748$ & $r(q_{89},q_{94})=0.672$ & $r(q_{89},q_{95})=0.860$ & $r(q_{90},q_{94})=-0.033$ \\
$r(q_{93},q_{94})=0.603$ & $r(q_{93},q_{95})=0.680$ & $r(q_{94},q_{95})=0.611$ &  &  \\
\mr
\multicolumn{5}{l}{Other miscellaneous correlation coefficients} \\
\mr
$r(q_{31},q_{57})\;\,=0.165$ & $r(q_{53},q_{68})\;\,=-0.672$ & $r(q_{81},q_{91})\;\,=-0.123$ & $r(q_{56},q_{106})=-0.277$ & $r(q_{58},q_{106})=-0.392$ \\
$r(q_{61},q_{106})=0.329$ & $r(q_{69},q_{106})=\lineup\m0.369$ & $r(q_{95},q_{106})=\lineup\m0.221$ &  &  \\
\br
\end{tabular}
\item[] $^a$ This subset of correlation coefficients was computed by D. B. Hume.
\item[] $^b$ This subset of correlation coefficients was provided by M. Pizzocaro. After approval of the 2021 recommended frequency values by the CCTF, an error in the calculation of $r(q_{24},q_{81})$ was identified. However using the corrected value of 0.155 does not change the results at the relevant level of precision.
\item[] $^c$ This subset of correlation coefficients was computed by N. Nemitz.
\item[] $^d$ This subset of correlation coefficients was computed by T. Kobayashi.
\end{indented}
\end{table*}

\subsection{Measurements involving PTB $^{171}$Yb$^+$ and $^{87}$Sr optical clocks}
The largest correlation coefficient in the input data set is between $q_{19}$~\cite{Huntemann2014} and $q_{45}$~\cite{Falke2014}. These two absolute frequency measurements, of the $^{171}$Yb$^+$ E2 and the $^{87}$Sr optical clock transitions, were performed at the same time, and the uncertainty arising from the local Cs fountains PTB-CSF1 and PTB-CSF2 completely dominates over the systematic uncertainty of either optical clock, resulting in a correlation coefficient $r(q_{19},q_{45})=0.981$~\cite{ROCIT_Guidelines}. Other much smaller correlations exist between another set of measurements performed with the PTB $^{171}$Yb$^+$ and $^{87}$Sr optical clocks ($q_{73}$, $q_{92}$, $q_{98}$ and $q_{99}$~\cite{Schwarz2020,Dorscher2021,Lange2021}), as the systematic uncertainty budgets of the clocks were common to the different measurements.

\subsection{Measurements involving $^{171}$Yb$^+$ trapped ion clocks at NPL and PTB}
\label{sec:Ybion_NPL_PTB}
Measurements $q_{14}$, $q_{18}$ and $q_{63}$ were all performed in the same campaign involving the NPL $^{171}$Yb$^+$ trapped ion optical clock~\cite{Godun2014}. The three correlation coefficients between these measurements are the only correlation coefficients that were included in the 2017 least-squares adjustment, and have contributions from both the statistical and systematic uncertainties of the clocks involved~\cite{ROCIT_Guidelines}. These measurements are also (less significantly) correlated with measurements $q_{13}$, $q_{16}$ and $q_{19}$ made at PTB~\cite{Tamm2014,Huntemann2012,Huntemann2014} because the blackbody radiation shifts for the NPL $^{171}$Yb$^+$ clock were calculated using experimental values for the differential polarizabilities of the atomic states determined at PTB~\cite{ROCIT_Guidelines}. 

\subsection{Measurements performed in June 2015}
Fourteen measurements in the input data set originate from June 2015, having been performed during a coordinated European clock comparison campaign. Five of these ($q_7$, $q_{47}$, $q_{49}$, $q_{58}$ and $q_{71}$) are absolute frequency measurements of $^{199}$Hg~\cite{Tyumenev2016}, $^{87}$Sr~\cite{Lodewyck2016, Grebing2016}, $^{87}$Rb~\cite{Guena2017} and $^{171}$Yb$^+$ (E3)~\cite{Baynham2018} atomic clocks. Three ($q_{60}$, $q_{61}$ and $q_{69}$) are local frequency ratio measurements between clocks at LNE-SYRTE~\cite{Tyumenev2016,Lodewyck2016}, and the remainder ($q_{84}$--$q_{87}$, $q_{100}$ and $q_{101}$) are remote optical frequency ratio measurements between $^{171}$Yb$^+$ (E3) and $^{87}$Sr optical clocks at NPL, LNE-SYRTE and PTB~\cite{Riedel2020}. Correlations were largely considered to be due to systematic uncertainties that were common to different measurements, with the largest coming from the Rb fountain at LNE-SYRTE, closely followed by the their $^{199}$Hg optical clock~\cite{ROCIT_Guidelines}. However contributions from statistical uncertainties were also estimated and included in some cases. 

\subsection{Optical frequency ratio measurements at NIST/JILA}
The frequency ratio measurements $q_{102}$, $q_{103}$ and $q_{104}$, which have the lowest uncertainties in the 2021 input data set, were performed using the $^{27}$Al$^+$, $^{171}$Yb and $^{87}$Sr optical clocks at NIST and JILA.
Correlations arise between pairs of ratios ($^{27}$Al$^+$/$^{171}$Yb, $^{171}$Yb/$^{87}$Sr and $^{27}$Al$^+$/$^{87}$Sr) due to the common atomic species and overlapping data.  Correlations due to both systematic uncertainty and statistical noise were considered.  The correlation due to statistical noise depends on the instability of each clock individually and the fraction of overlapping data.  While the exact instability of each clock is unknown, the instabilities of the lattice clocks contribute negligibly to the correlation coefficients and were assumed to be equal.  On the other hand, the $^{27}$Al$^+$ clock instability dominates the statistical noise for the $^{27}$Al$^+$ ratios so it can be determined from the measurements and contributes significantly to the correlation coefficient for those ratios.  There is additional correlation due to fluctuations from the density shift in $^{87}$Sr (evaluated daily) and the observed excess scatter in the $^{27}$Al$^+$/$^{171}$Yb and $^{171}$Yb/$^{87}$Sr ratios.  Both were included as additional uncertainties acting day-to-day.  The level of excess scatter was chosen to make the reduced chi-squared value equal to 1 for both $^{27}$Al$^+$/$^{171}$Yb and $^{171}$Yb/$^{87}$Sr and is consistent with the histograms determined from Bayesian analysis in~\cite{BACON2021}.

\subsection{Measurements involving the INRIM $^{171}$Yb lattice clock}
Measurements $q_{24}$, $q_{76}$, $q_{81}$, $q_{82}$ and $q_{83}$ all involve the INRIM $^{171}$Yb lattice clock. 
Measurements $q_{76}$, $q_{82}$ and $q_{83}$ were obtained in the same period of time, and are therefore correlated through the systematic uncertainty of the $^{171}$Yb clock, as well as through extrapolations of maser noise over common periods of dead time. 
Measurements $q_{82}$ and $q_{83}$ are both measurements against the NICT $^{87}$Sr optical lattice clock, and hence correlated through the systematic uncertainty of that clock. The systematic uncertainty of the INRIM $^{171}$Yb lattice clock was re-evaluated in 2019 so there are no significant correlations between earlier and later measurements.

\subsection{Measurements involving optical clocks at RIKEN}
Frequency ratio measurements $q_{59}$, $q_{66}$ and $q_{79}$~\cite{Yamanaka2015,Nemitz2016,Ohmae2020} involve the $^{87}$Sr, $^{171}$Yb and $^{199}$Hg optical clocks at RIKEN. 
The correlation coefficient between the two measurements involving the $^{199}$Hg optical lattice clock is particularly significant, with a value of 0.826, since the evaluation of systematic frequency shifts and uncertainty were determined in an identical evaluation campaign.

\subsection{Measurements involving the $^{115}$In$^+$ and $^{87}$Sr optical clocks at NICT}
\label{sec:NICT_Inion_Sr}
The measurements $q_{3}$, $q_{41}$, $q_{50}$, $q_{74}$, $q_{78}$, $q_{80}$ and $q_{90}$~\cite{Ohtsubo2017,Yamaguchi2012,Hachisu2017a,Ohtsubo2020,Fujieda2018,Nemitz2021} involve the $^{115}$In$^+$ and $^{87}$Sr optical clocks at NICT. The largest correlation coefficient by far is between $q_{74}$ (an absolute frequency measurement of the $^{115}$In$^+$ clock transition) and $q_{78}$ (an $^{115}$In$^+$ / $^{87}$Sr optical frequency ratio measurement), with a value of $r(q_{74}, q_{78})=0.859$. These two measurements were performed during the same campaign, with the uncertainty being dominated by that of the $^{115}$In$^+$ optical clock (both systematics and statistics). 

\subsection{Measurements involving the $^{171}$Yb and $^{87}$Sr optical lattice clocks at NMIJ}
Measurements $q_{89}$, $q_{93}$, $q_{94}$ and $q_{95}$ involve the $^{171}$Yb and $^{87}$Sr optical lattice clocks at NMIJ. The optical frequency ratio between the two clocks, $q_{93}$~\cite{Hisai2021} was measured during the last part of the campaign in which the absolute frequency $q_{89}$ was measured~\cite{Kobayashi2020}, resulting in a significant correlation coefficient $r(q_{89}, q_{93})=0.748$, arising mainly from the systematic uncertainty of the $^{171}$Yb lattice clock. The absolute frequency measurement was made via a comparison to TAI, during which period the NICT $^{87}$Sr optical clock and the SYRTE $^{87}$Rb fountain were amongst the standards used to calibrate TAI. This meant that it was possible to determine the $^{171}$Yb / $^{87}$Sr ($q_{94}$) and $^{171}$Yb / $^{87}$Rb ratios ($q_{95}$) in the same campaign. However the correlation coefficients 
$r(q_{89}, q_{94})$, $r(q_{89}, q_{95})$, $r(q_{93}, q_{94})$ and $r(q_{93}, q_{95})$ are also large, ranging from 0.603 to 0.860. 
Measurement $q_{94}$ is correlated with $q_{78}$ and $q_{90}$ since NICT-Sr1 is a common standard. However these correlation coefficients are less significant due to a larger link uncertainty between Yb and Sr.

\subsection{Miscellaneous correlation coefficients}
Absolute frequency measurements $q_{31}$~\cite{Barwood2014} and $q_{57}$~\cite{Ovchinnikov2015} were performed in a partly common period against the local caesium fountain NPL-CsF1, and hence are correlated, mainly through the systematic uncertainty of the fountain. The two measurements $q_{53}$ and $q_{68}$ reported in~\cite{Matsubara2012}, are correlated through the common systematic uncertainty of the NICT $^{40}$Ca$^+$ optical clock. 
Measurements $q_{91}$ and $q_{81}$, both reported in~\cite{Grotti2018}, are correlated because the systematic uncertainty of the PTB transportable $^{87}$Sr optical lattice clock is common to the two measurements.
The $^{171}$Yb / $^{87}$Rb ratio $q_{106}$ was determined through measurements made against TAI, and hence is correlated with several other measurements ($q_{56}$, $q_{58}$, $q_{61}$, $q_{69}$ and $q_{95}$) involving the LNE-SYRTE Rb fountain.

\section{Frequency ratios}
\label{sec:OutputRatios}

Table~\ref{tab:Output_Ratios} lists the frequency ratio values consistent with the 2021 recommended frequency values, taking into account correlations between those values. Ratios involving $\nu_2$ and $\nu_9$ ($^1$H and $^{40}$Ca) are excluded from the list, as these two recommended frequency values are not correlated with the others. 

\begin{table*}
\caption{\label{tab:Output_Ratios}Frequency ratios consistent with the 2021 recommended frequency values, taking into account the covariance of the output matrix. ($\nu_2$ and $\nu_9$ are linked to the other frequencies only via Cs ($\nu_{15}$), and hence are not included in this table.)} 
\begin{indented}
\item[]\begin{tabular}{@{}llll}
\br
Clock transitions & Atomic species  &  Frequency ratio & Fractional uncertainty \\
\mr
$\nu_1/\nu_3$    & $^{115}$In$^+$/$\,^{199}$Hg          & {\lineup\0\0\0\0\0\,}1.123\,010\,988\,476\,8743(49)  & $4.3\times 10^{-15}$ \\
$\nu_1/\nu_4$    & $^{115}$In$^+$/$\,^{27}$Al$^+$       & {\lineup\0\0\0\0\0\,}1.130\,584\,343\,961\,8487(49)  & $4.3\times 10^{-15}$ \\
$\nu_1/\nu_5$    & $^{115}$In$^+$/$\,^{199}$Hg$^+$      & {\lineup\0\0\0\0\0\,}1.190\,360\,410\,756\,6604(51)  & $4.3\times 10^{-15}$ \\
$\nu_1/\nu_6$    & $^{115}$In$^+$/$\,^{171}$Yb$^+$(E2)  & {\lineup\0\0\0\0\0\,}1.841\,194\,044\,091\,2659(80)  & $4.3\times 10^{-15}$ \\
$\nu_1/\nu_7$    & $^{115}$In$^+$/$\,^{171}$Yb$^+$(E3)  & {\lineup\0\0\0\0\0\,}1.973\,773\,591\,557\,2195(85)  & $4.3\times 10^{-15}$ \\
$\nu_1/\nu_8$    & $^{115}$In$^+$/$\,^{171}$Yb          & {\lineup\0\0\0\0\0\,}2.445\,326\,324\,126\,955(11) & $4.3\times 10^{-15}$ \\
$\nu_1/\nu_{10}$ & $^{115}$In$^+$/$\,^{88}$Sr$^+$       & {\lineup\0\0\0\0\0\,}2.849\,510\,267\,459\,795(13) & $4.5\times 10^{-15}$ \\
$\nu_1/\nu_{11}$ & $^{115}$In$^+$/$\,^{88}$Sr           & {\lineup\0\0\0\0\0\,}2.952\,748\,322\,069\,815(13) & $4.3\times 10^{-15}$ \\
$\nu_1/\nu_{12}$ & $^{115}$In$^+$/$\,^{87}$Sr           & {\lineup\0\0\0\0\0\,}2.952\,748\,749\,874\,866(13) & $4.3\times 10^{-15}$ \\
$\nu_1/\nu_{13}$ & $^{115}$In$^+$/$\,^{40}$Ca$^+$       & {\lineup\0\0\0\0\0\,}3.083\,388\,200\,597\,554(14) & $4.7\times 10^{-15}$ \\
$\nu_1/\nu_{14}$ & $^{115}$In$^+$/$\,^{87}$Rb           & 185\,436.914\,199\,787\,30(80) & $4.3\times 10^{-15}$ \\
\ms
$\nu_3/\nu_4$    & $^{199}$Hg$\,$/$\,^{27}$Al$^+$      & {\lineup\0\0\0\0\0\,}1.006\,743\,794\,640\,198\,49(15) & $1.5\times 10^{-16}$ \\
$\nu_3/\nu_5$    & $^{199}$Hg$\,$/$\,^{199}$Hg$^+$     & {\lineup\0\0\0\0\0\,}1.059\,972\,184\,574\,196\,57(19) & $1.8\times 10^{-16}$ \\
$\nu_3/\nu_6$    & $^{199}$Hg$\,$/$\,^{171}$Yb$^+$(E2) & {\lineup\0\0\0\0\0\,}1.639\,515\,608\,470\,095\,42(28) & $1.7\times 10^{-16}$ \\
$\nu_3/\nu_7$    & $^{199}$Hg$\,$/$\,^{171}$Yb$^+$(E3) & {\lineup\0\0\0\0\0\,}1.757\,572\,821\,468\,313\,31(27) & $1.5\times 10^{-16}$ \\
$\nu_3/\nu_8$    & $^{199}$Hg$\,$/$\,^{171}$Yb         & {\lineup\0\0\0\0\0\,}2.177\,473\,194\,134\,564\,88(32) & $1.5\times 10^{-16}$ \\
$\nu_3/\nu_{10}$ & $^{199}$Hg$\,$/$\,^{88}$Sr$^+$      & {\lineup\0\0\0\0\0\,}2.537\,384\,136\,663\,3019(34) & $1.4\times 10^{-15}$ \\
$\nu_3/\nu_{11}$ & $^{199}$Hg$\,$/$\,^{88}$Sr          & {\lineup\0\0\0\0\0\,}2.629\,313\,828\,954\,238\,79(40) & $1.5\times 10^{-16}$ \\
$\nu_3/\nu_{12}$ & $^{199}$Hg$\,$/$\,^{87}$Sr          & {\lineup\0\0\0\0\0\,}2.629\,314\,209\,898\,909\,56(39) & $1.5\times 10^{-16}$ \\
$\nu_3/\nu_{13}$ & $^{199}$Hg$\,$/$\,^{40}$Ca$^+$      & {\lineup\0\0\0\0\0\,}2.745\,643\,838\,071\,0009(49) & $1.8\times 10^{-15}$ \\
$\nu_3/\nu_{14}$ & $^{199}$Hg$\,$/$\,^{87}$Rb          & 165\,124.754\,879\,997\,262(60) & $3.6\times 10^{-16}$ \\
\ms
$\nu_4/\nu_5$    & $^{27}$Al$^+$/$\,^{199}$Hg$^+$     & {\lineup\0\0\0\0\0\,}1.052\,871\,833\,148\,990\,45(11) & $1.0\times 10^{-16}$ \\
$\nu_4/\nu_6$    & $^{27}$Al$^+$/$\,^{171}$Yb$^+$(E2) & {\lineup\0\0\0\0\0\,}1.628\,533\,115\,573\,902\,39(14) & $8.3\times 10^{-17}$ \\
$\nu_4/\nu_7$    & $^{27}$Al$^+$/$\,^{171}$Yb$^+$(E3) & {\lineup\0\0\0\0\0\,}1.745\,799\,508\,102\,709\,104(84) & $4.8\times 10^{-17}$ \\
$\nu_4/\nu_8$    & $^{27}$Al$^+$/$\,^{171}$Yb         & {\lineup\0\0\0\0\0\,}2.162\,887\,127\,516\,663\,705(24)  & $1.1\times 10^{-17}$\\
$\nu_4/\nu_{10}$ & $^{27}$Al$^+$/$\,^{88}$Sr$^+$      & {\lineup\0\0\0\0\0\,}2.520\,387\,163\,220\,7488(34) & $1.3\times 10^{-15}$ \\
$\nu_4/\nu_{11}$ & $^{27}$Al$^+$/$\,^{88}$Sr          & {\lineup\0\0\0\0\0\,}2.611\,701\,053\,388\,596\,03(10) & $3.9\times 10^{-17}$ \\
$\nu_4/\nu_{12}$ & $^{27}$Al$^+$/$\,^{87}$Sr          & {\lineup\0\0\0\0\0\,}2.611\,701\,431\,781\,463\,019(39) & $1.5\times 10^{-17}$ \\
$\nu_4/\nu_{13}$ & $^{27}$Al$^+$/$\,^{40}$Ca$^+$      & {\lineup\0\0\0\0\0\,}2.727\,251\,811\,919\,3078(48) & $1.8\times 10^{-15}$ \\
$\nu_4/\nu_{14}$ & $^{27}$Al$^+$/$\,^{87}$Rb          & 164\,018.646\,808\,755\,766(54) & $3.3\times 10^{-16}$ \\
\ms
$\nu_5/\nu_6$    & $^{199}$Hg$^+$/$\,^{171}$Yb$^+$(E2) & {\lineup\0\0\0\0\0\,}1.546\,753\,426\,486\,100\,05(21) & $1.3\times 10^{-16}$ \\
$\nu_5/\nu_7$    & $^{199}$Hg$^+$/$\,^{171}$Yb$^+$(E3) & {\lineup\0\0\0\0\0\,}1.658\,131\,078\,387\,072\,22(19) & $1.1\times 10^{-16}$ \\
$\nu_5/\nu_8$    & $^{199}$Hg$^+$/$\,^{171}$Yb         & {\lineup\0\0\0\0\0\,}2.054\,273\,900\,601\,723\,59(22) & $1.0\times 10^{-16}$ \\
$\nu_5/\nu_{10}$ & $^{199}$Hg$^+$/$\,^{88}$Sr$^+$      & {\lineup\0\0\0\0\0\,}2.393\,821\,435\,684\,7480(32) & $1.3\times 10^{-15}$ \\
$\nu_5/\nu_{11}$ & $^{199}$Hg$^+$/$\,^{88}$Sr          & {\lineup\0\0\0\0\0\,}2.480\,549\,836\,324\,681\,89(28) & $1.1\times 10^{-16}$ \\
$\nu_5/\nu_{12}$ & $^{199}$Hg$^+$/$\,^{87}$Sr          & {\lineup\0\0\0\0\0\,}2.480\,550\,195\,715\,877\,54(26) & $1.1\times 10^{-16}$ \\
$\nu_5/\nu_{13}$ & $^{199}$Hg$^+$/$\,^{40}$Ca$^+$      & {\lineup\0\0\0\0\0\,}2.590\,298\,007\,842\,4970(46) & $1.8\times 10^{-15}$ \\
$\nu_5/\nu_{14}$ & $^{199}$Hg$^+$/$\,^{87}$Rb          & 155\,782.158\,516\,102\,797(54) & $3.5\times 10^{-16}$ \\
\ms
$\nu_6/\nu_7$    & $^{171}$Yb$^+$(E2)$\,$/$\,^{171}$Yb$^+$(E3) & {\lineup\0\0\0\0\0\,}1.072\,007\,373\,634\,205\,473(73) & $6.9\times 10^{-17}$ \\
$\nu_6/\nu_8$    & $^{171}$Yb$^+$(E2)$\,$/$\,^{171}$Yb         & {\lineup\0\0\0\0\0\,}1.328\,119\,831\,787\,671\,42(11)  & $8.3\times 10^{-17}$ \\
$\nu_6/\nu_{10}$ & $^{171}$Yb$^+$(E2)$\,$/$\,^{88}$Sr$^+$      & {\lineup\0\0\0\0\0\,}1.547\,642\,561\,958\,3136(21)    & $1.3\times 10^{-15}$ \\
$\nu_6/\nu_{11}$ & $^{171}$Yb$^+$(E2)$\,$/$\,^{88}$Sr          & {\lineup\0\0\0\0\0\,}1.603\,713\,813\,623\,139\,52(14) & $8.9\times 10^{-17}$ \\
$\nu_6/\nu_{12}$ & $^{171}$Yb$^+$(E2)$\,$/$\,^{87}$Sr          & {\lineup\0\0\0\0\0\,}1.603\,714\,045\,975\,103\,00(13) & $8.2\times 10^{-17}$ \\
$\nu_6/\nu_{13}$ & $^{171}$Yb$^+$(E2)$\,$/$\,^{40}$Ca$^+$      & {\lineup\0\0\0\0\0\,}1.674\,667\,703\,001\,0606(30)   & $1.8\times 10^{-15}$  \\
$\nu_6/\nu_{14}$ & $^{171}$Yb$^+$(E2)$\,$/$\,^{87}$Rb          & 100\,715.573\,567\,538\,329(34)  & $3.4\times 10^{-16}$  \\
\ms
$\nu_7/\nu_8$    & $^{171}$Yb$^+$(E3)$\,$/$\,^{171}$Yb    & {\lineup\0\0\0\0\0\,}1.238\,909\,231\,832\,259\,428(59) & $4.7\times 10^{-17}$ \\
$\nu_7/\nu_{10}$ & $^{171}$Yb$^+$(E3)$\,$/$\,^{88}$Sr$^+$ & {\lineup\0\0\0\0\0\,}1.443\,686\,489\,498\,3514(19) & $1.3\times 10^{-15}$ \\
$\nu_7/\nu_{11}$ & $^{171}$Yb$^+$(E3)$\,$/$\,^{88}$Sr     & {\lineup\0\0\0\0\0\,}1.495\,991\,401\,800\,156\,824(86) & $5.8\times 10^{-17}$ \\
$\nu_7/\nu_{12}$ & $^{171}$Yb$^+$(E3)$\,$/$\,^{87}$Sr     & {\lineup\0\0\0\0\0\,}1.495\,991\,618\,544\,900\,552(68) & $4.6\times 10^{-17}$ \\
$\nu_7/\nu_{13}$ & $^{171}$Yb$^+$(E3)$\,$/$\,^{40}$Ca$^+$ & {\lineup\0\0\0\0\0\,}1.562\,179\,276\,177\,7189(28) & $1.8\times 10^{-15}$ \\
$\nu_7/\nu_{14}$ & $^{171}$Yb$^+$(E3)$\,$/$\,^{87}$Rb     & {\lineup\0}93\,950.448\,518\,001\,415(31) & $3.3\times 10^{-16}$ \\
\ms
$\nu_8/\nu_{10}$ & $^{171}$Yb$\,$/$\,^{88}$Sr$^+$ & {\lineup\0\0\0\0\0\,}1.165\,288\,345\,913\,1553(16) & $1.3\times 10^{-15}$ \\
$\nu_8/\nu_{11}$ & $^{171}$Yb$\,$/$\,^{88}$Sr     & {\lineup\0\0\0\0\0\,}1.207\,506\,864\,395\,296\,327(46) & $3.8\times 10^{-17}$ \\
$\nu_8/\nu_{12}$ & $^{171}$Yb$\,$/$\,^{87}$Sr     & {\lineup\0\0\0\0\0\,}1.207\,507\,039\,343\,337\,845(16) & $1.3\times 10^{-17}$ \\
$\nu_8/\nu_{13}$ & $^{171}$Yb$\,$/$\,^{40}$Ca$^+$ & {\lineup\0\0\0\0\0\,}1.260\,931\,177\,231\,8993(22) & $1.8\times 10^{-15}$ \\
$\nu_8/\nu_{14}$ & $^{171}$Yb$\,$/$\,^{87}$Rb     & {\lineup\0}75\,833.197\,545\,114\,200(25) & $3.3\times 10^{-16}$ \\
\ms
$\nu_{10}/\nu_{11}$ & $^{88}$Sr$^+$/$\,^{88}$Sr     & {\lineup\0\0\0\0\0\,}1.036\,230\,104\,446\,0007(14) & $1.3\times 10^{-15}$ \\
$\nu_{10}/\nu_{12}$ & $^{88}$Sr$^+$/$\,^{87}$Sr     & {\lineup\0\0\0\0\0\,}1.036\,230\,254\,578\,8345(14) & $1.3\times 10^{-15}$ \\
$\nu_{10}/\nu_{13}$ & $^{88}$Sr$^+$/$\,^{40}$Ca$^+$ & {\lineup\0\0\0\0\0\,}1.082\,076\,536\,381\,8990(24) & $2.2\times 10^{-15}$ \\
$\nu_{10}/\nu_{14}$ & $^{88}$Sr$^+$/$\,^{87}$Rb     & {\lineup\0}65\,076.766\,459\,625\,929(88) & $1.4\times 10^{-15}$ \\
\ms
$\nu_{11}/\nu_{12}$ & $^{88}$Sr$\,$/$\,^{87}$Sr     & {\lineup\0\0\0\0\0\,}1.000\,000\,144\,883\,682\,799(36) & $3.6\times 10^{-17}$ \\
$\nu_{11}/\nu_{13}$ & $^{88}$Sr$\,$/$\,^{40}$Ca$^+$ & {\lineup\0\0\0\0\0\,}1.044\,243\,485\,823\,4592(18) & $1.8\times 10^{-15}$ \\
$\nu_{11}/\nu_{14}$ & $^{88}$Sr$\,$/$\,^{87}$Rb     & {\lineup\0}62\,801.462\,899\,418\,361(21) & $3.3\times 10^{-16}$ \\
& & \\
\ms
$\nu_{12}/\nu_{13}$ & $^{87}$Sr$\,$/$\,^{40}$Ca$^+$ & {\lineup\0\0\0\0\0\,}1.044\,243\,334\,529\,6392(18) & $1.8\times 10^{-15}$ \\
$\nu_{12}/\nu_{14}$ & $^{87}$Sr$\,$/$\,^{87}$Rb     & {\lineup\0}62\,801.453\,800\,512\,449(21) & $3.3\times 10^{-16}$ \\
\ms
$\nu_{13}/\nu_{14}$ & $^{40}$Ca$^+$/$\,^{87}$Rb & {\lineup\0}60~140.631~712~818~40(11) & $1.8\times 10^{-15}$ \\
\br
\end{tabular}
\end{indented}
\end{table*}

\section*{References} 

\bibliography{2021RecValues}

\end{document}